\documentclass[12pt]{article}
\usepackage{epsfig,amsfonts,amssymb}
\usepackage{hyperref}
\usepackage{cite}
\input epsf.sty
\def\ZZZ{{\hbox{ Z\kern-1.6mm Z}}}
\def\zzz{{\hbox{z\kern-1mm z}}}

\newcommand{\vt}{\vartheta}

\def\RRR{{\hbox{ R\kern-2.4mm R}}}

\newcommand{\JJ}{{\cal J}}

\newcommand{\OO}{{\cal O}}

\newcommand{\EE}{{\cal E}}
\newcommand{\LL}{{\cal L}}

\newcommand{\wt}{\widetilde}
\newcommand{\wh}{\widehat}

\newcommand{\NN}{{\cal N}}

\newcommand{\be}{\begin{equation}}
\newcommand{\ee}{\end{equation}}
\newcommand{\ben}{\begin{eqnarray}\displaystyle}
\newcommand{\een}{\end{eqnarray}}

\newcommand{\bea}[1]{\begin{eqnarray}\label{#1} }
\newcommand{\eea}{\end{eqnarray}}

\newcommand{\refb}[1]{(\ref{#1})}
\newcommand{\p}{\partial}
\newcommand{\sectiono}[1]{\section{#1}\setcounter{equation}{0}}

\def\one{{\hbox{ 1\kern-.8mm l}}}
\def\zero{{\hbox{ 0\kern-1.5mm 0}}}

\begin{document}

\begin{flushright}
arXiv:1008.3801
\end{flushright}

\baselineskip 24pt

\begin{center} 

{\Large \bf Black Hole Microstate 
Counting and its Macroscopic Counterpart}

\end{center}

\vskip .6cm
\medskip

\vspace*{4.0ex}

\baselineskip=18pt

\centerline{\large \rm Ipsita Mandal$^a$ and 
Ashoke Sen$^{a,b}$}

\vspace*{4.0ex}
\centerline{\large \it $^a$Harish-Chandra Research Institute}
\centerline{\large \it  Chhatnag Road, Jhusi,
Allahabad 211019, India}
\vspace*{3.0ex}
\centerline{\large \it 
$^b$LPTHE, Universite Pierre et Marie Curie, Paris 6}
\centerline{\large \it 
4 Place Jussieu,  75252 Paris Cedex 05, France}

\vspace*{1.0ex}
\centerline{E-mail:  ipsita, sen at hri.res.in}

\vspace*{5.0ex}

\centerline{\bf Abstract} \bigskip

We survey recent results on the exact dyon spectrum in a class
of $\NN=4$ supersymmetric string theories, and discuss how
the results can be understood from the macroscopic viewpoint
using $AdS_2/CFT_1$ correspondence.
The comparison
between the microscopic and the macroscopic results includes 
power suppressed corrections
to the entropy, the
sign of the index, logarithmic corrections 
and also the twisted index measuring the
distribution of discrete quantum numbers among the microstates.

(Based on lectures given by A.S. at the
12th Marcel Grossmann Meeting On General Relativity,
12-18 Jul 2009, Paris, France;
CERN Winter School on Supergravity, 
Strings, and Gauge Theory,
 25-29 January 2010; String Theory: Formal Developments And Applications,
21 Jun - 3 Jul 2010, Cargese, France, and notes taken
by I.M. at the Cargese school.)

\vfill \eject

\baselineskip=18pt

\tableofcontents

\renewcommand{\theequation}{\thesection.\arabic{equation}}

\sectiono{Introduction} \label{s00}

Black holes are  classical solutions of the equations of 
motion of general theory of relativity.
Each black hole is surrounded by an event horizon that acts as
a one way membrane.
Nothing, including light, can escape a black hole horizon.
Thus classically the
horizon of a black hole behaves as 
a perfect black body at zero temperature. 

This picture undergoes a dramatic modification in quantum
theory \cite{r1,r2,r3,r4}.  
There
a black hole behaves as a thermodynamic system with definite
temperature, entropy etc. 
In particular, the temperature
and the Bekenstein-Hawking entropy of a black hole is given by the
simple formul$\ae$:
\be \label{eip1}
T = {\kappa  \over 2\pi }, \qquad S_{BH} = 
{A \over 4 \, G_N },
\ee
where $\kappa$ is the surface gravity --
acceleration due to gravity at the horizon of the
black hole (measured by an observer at infinity),   
$ A $ is the area of the event horizon and 
$ G_N $ is the Newton's gravitational constant.  
We have set $ \hbar = c =k_B = 1$.

Now, for ordinary objects, the entropy of a system has a
microscopic interpretation. If we fix the macroscopic 
parameters ({\it e.g.} total electric charge, energy etc.)
and count the number of quantum states (dubbed microstates), 
each of which has the same charge, energy etc., then we can 
define the microscopic (statistical) entropy as:
\be \label{ei2}
S_{micro} = \ln \, d_{micro} \,,
\ee
where $d_{micro}$ is the number of such microstates.
This naturally leads to the question 
whether the entropy of a black hole has a similar 
statistical interpretation. As pointed out by Hawking, 
answering this question in the affirmative is essential for any
consistent theory of quantum gravity as otherwise 
it leads to violation of the laws of quantum mechanics.

In order to investigate the statistical origin 
of black hole entropy, we need a quantum 
theory of gravity. Since string theory gives a framework 
for studying classical and quantum properties of black holes,
we shall carry out our investigation in string theory.
Now, even though there is a unique string (M)-theory, 
it can exist in many different
stable and metastable phases. Without knowing precisely 
which phase of string theory
describes the part of the universe we live in, we cannot directly
compare string theory to experiments. However, there 
are some issues like those involving black hole thermodynamics, 
which are universal,
 and hence can be addressed in any phase of string theory.
 We shall make use of this freedom to study these
 issues in a special class of phases of string theory with
 a large amount of unbroken supersymmetry. 
Since these phases
have Bose-Fermi degenerate spectrum of states,
they do not describe the observed world. Nevertheless they
contain black hole solutions and hence can be used to study
issues involving black hole thermodynamics. 

Many aspects of black hole thermodynamics 
have been studied in
string theory, but we shall 
focus our
attention on one particular aspect: entropy of the black 
hole in the zero temperature 
limit (\i.e., supersymmetric, extremal black holes). The 
advantage of studying such a black hole is that it is a 
stable state of the theory.
The general strategy is as follows\cite{9504147,9601029}:
\begin{enumerate}
\item Identify a supersymmetric black hole carrying a certain set
of electric charges $\{Q_i\}$ and magnetic charges $\{P_i\}$, and
calculate its entropy $S_{BH}(Q,P)$ using the Bekenstein-Hawking
formula.\footnote{Since we are considering a 
generic phase of string theory, it may
have more that one Maxwell field and hence multiple charges.}
\item Identify the supersymmetric
quantum states in string theory carrying the same set
of charges. These can include not only the fundamental 
strings but also other objects
in string theory which are required for consistency 
of the theory ({\it e.g.} D-branes, Kaluza-Klein monopoles).
We then calculate the number 
$d_{micro}(Q,P)$ of these states.
\item Compare $ S_{micro}\equiv 
\ln\, d_{micro}(Q,P)$ with $S_{BH}(Q,P)$.
\end{enumerate}

For a class of supersymmetric
extremal black holes in type IIB string theory on $K3\times S^1$,
 Strominger and
Vafa\cite{9601029} 
computed the Bekenstein-Hawking entropy via 
\refb{eip1}
and found agreement with the statistical entropy 
defined in \refb{ei2}. This agreement is quite remarkable 
since it relates a geometric
quantity in black hole space-time to a counting problem 
that does not make
any direct reference to black holes.
At the same time, one should keep in mind that
the Bekenstein-Hawking formula is an
approximate formula that holds in classical general theory
of relativity. While string theory gives a theory 
of gravity that reduces to
Einstein's theory when gravity is weak, there are 
corrections.\footnote{In string theory, even at 
classical level, we have higher derivative ($\alpha'$) 
corrections. This is because strings are not point objects. 
So even at classical level, there will be corrections to the  
Bekenstein-Hawking formula. Besides this, there will also
be quantum corrections.} 
Thus the Bekenstein-Hawking formula for the entropy works
well only when gravity at the horizon is weak. 
Typically this requires the charges to be large. 
Similarly,
the computation of $d_{micro}$ in \cite{9601029} was 
also carried out
in the limit of large charges,  so that instead of having to carry out an
exact counting of states, one can use some appropriate
asymptotic formula to compute it. 
Thus the agreement between $S_{BH}$
and $S_{micro}$, seen in \cite{9601029}, can be regarded as
an agreement in the limit of large size.

This leads to the following question:
For ordinary systems, thermodynamics 
provides an accurate description only      
in the limit of large volume. 
Is the situation with black holes similar, \i.e.,  do they only
capture the information about the system in the limit of large
charge and mass? Or, could it be that the 
relation $ A/ 4 G_N = \ln   d_{micro}$ is
an approximation to an exact result? Our goal 
will be to argue for the second possibility by giving
an exact formula to which the
above is an approximation.

In order to address this issue, we have to work on two fronts:
\begin{enumerate}
\item Count the number of microstates to greater accuracy.
\item Calculate the black hole entropy to greater accuracy.
\end{enumerate}
We can then compare the two to see if they agree beyond the large
charge limit. In these lectures we shall 
describe the progress on both fronts.

Note that on the gravity side we shall not try to identify
the individual microstates -- this is the goal of the fuzzball 
program \cite{0810.4525}. 
Our approach will be to find a systematic
procedure that allows us to compute the total number of
states in the ensemble from the gravity 
side without having to identify the individual
microstates. More
generally, we would like to find an algorithm for
computing the trace of various observables in
this ensemble from the gravity side.

We end this section by giving a summary of the progress, which will
be reviewed in detail in the rest of these lecture notes:
\begin{enumerate}
{\bf \item 
Progress in microscopic counting:}
In a wide class of phases of string theory
with 16 or more unbroken supercharges, one 
now has a complete understanding of the
microscopic `degeneracies' of supersymmetric 
black holes \cite{9607026,0412287,0505094,0506151,0506249,0508174,
0510147,
0602254,0603066,0605210,0607155,0609109,0612011,
0702141,0702150,0705.1433,0705.3874,0706.2363,0707.1563,0707.3035,
0708.1270,0708.3715,0710.4533,0712.0043,
0801.0149,0802.0544,0802.0761,0802.1556,0803.1014,
0803.2692,0803.3857,0806.2337,0807.0237,0807.1314,
0807.4451,0808.1746,0809.1157,0809.4258,0810.3472,
0901.1758,0903.2481,0907.1410,
0911.0586,0911.1563,1002.3857,1006.3472}. 
Typically, such theories have multiple Maxwell fields 
and the black hole is characterized by multiple 
electric and magnetic charges, collectively denoted by 
$(Q,P)$. It turns out that for a wide class of charge vectors
(all charge vectors in some cases),
$d_{micro}(Q,P)$ in these theories can be explicitly
computed and can be expressed as Fourier expansion
coefficients of some  functions with remarkable
symmetry properties. 
This provides us with the `experimental data' 
to be explained by a `theory of black holes', giving
a powerful tool 
for checking the internal consistency of string theory.
Needless to say, 
in the large charge limit, these degeneracies agree with
the exponential of the Bekenstein-Hawking entropy of black
holes carrying the same set of charges. 
Our goal will be to see how far the agreement can be pushed
beyond the large charge limit.

{\bf \item Progress in black hole entropy computation:}
On the macroscopic side, we would like to 
ask whether we can find an exact formula 
for the black hole entropy that
can be compared with $\ln d_{micro}(Q,P)$. 
This will require us to take into account
\begin{enumerate}
\item stringy $(\alpha'$)
corrections, and
\item quantum ($g_s$) corrections.
\end{enumerate}
We shall describe an approach to finding such a general
formula for black hole entropy using $AdS_2/CFT_1$
correspondence. We shall then apply this general formalism to the
specific case of supersymmetric black holes
in $\NN=4$ supersymmetric string theories,
and compare the results with the
microscopic answer.
\end{enumerate}

\sectiono{Microstate counting} \label{stwo}

In this section we shall survey the known results on the
counting of quarter-BPS dyons in $\NN=4$ supersymmetric
string theories.

\subsection{The role of index}

The counting of microstates is always done in a region of the
moduli space where
gravity is weak and hence the states do not form  a black hole. 
In order to be able to compare it with the black hole entropy, 
we must focus on quantities which do not change as we change
the coupling from small to large value. So we need an appropriate 
index which is protected by supersymmetry, and at the same
time does not vanish identically when evaluated on the microstates
of interest. The relevant index 
in $D=4$ turns out to be the helicity trace index \cite{9611205,9708062}.

Suppose we have a BPS state that breaks $4n$ supersymmetries. Then 
there will be $4n$ fermion zero modes (goldstinos) on the world-line 
of the state. Quantization of these zero modes will produce 
Bose-Fermi degenerate states. Thus the usual Witten index
$Tr(-1)^F$, which measures the difference between the number
of bosonic and fermionic states, 
will receive vanishing contribution from these states.
To remedy this situation, we define a new index called the helicity
trace index:
\be \label{heltr}
B_{2n} = {1\over (2n)!}\, Tr\lbrace(-1)^F (2h)^{2n}\rbrace = 
{1\over (2n)!}\, Tr\lbrace(-1)^{2h} (2h)^{2n}\rbrace\,,
\ee
where $h$ is the third component of the 
angular momentum in the rest frame.
The trace is taken over states carrying 
a fixed set of charges.
For every pair of fermion zero modes, 
$Tr\lbrace(-1)^F (2h)\rbrace$ gives a
 non-vanishing result $i$, leading to a non-zero 
contribution $(-1)^n$ to $B_{2n}$. On the other
hand, any state that breaks more than 
$4n$ supersymmetries, will have more
then $2n$ pairs of fermion zero modes and 
will give vanishing contribution
to this trace. In particular, non-BPS states will not contribute,
and the index will be protected from corrections as we vary the
moduli (except at the walls of marginal 
stability \cite{0010222,0101135,0206072,0304094,0702146},
which will be discussed
in \S\ref{swall}).  

Quarter-BPS black holes in $\NN=4$ supersymmetric string theories
preserve four of the sixteen supersymmetries, and hence break
twelve supersymmetries. Thus the relevant helicity trace index
is $B_6$. 
We shall now describe the microscopic results for $B_6$ in a
class of $\NN=4$ supersymmetric string theories. However, we
must keep in mind that,
since on the microscopic side we compute an index,
on the black hole side also we must compute an index. 
Otherwise we cannot compare the results of microscopic
and macroscopic computations. We will show in 
\S\ref{sdeg} how can we use black hole
entropy to compute the index $B_6$ on 
the black hole side.

\subsection{Microstate counting
in heterotic string theory on $T^6$} 

The simplest example of an $\NN=4$ supersymmetric
string theory is heterotic string theory on $T^6$ (or
equivalently type IIA or IIB string theory on
$K3\times T^2$, as they are related by duality transformations). 
This theory has 28 U(1) gauge fields arising from the Cartan
generators of the $E_8\times E_8$ (or $SO(32)$) gauge
group, and the components of the metric and the 2-form field
along the six internal directions. Thus a generic charged state 
is characterized  by 28 dimensional
electric charge vector $Q$ and 28 dimensional
magnetic charge vector $ P$. Under the
$ O(6,22;\ZZZ)$ T-duality symmetry of the theory, the
charges
$Q$ and $P$ transform as vectors. This allows us to define 
T-duality invariant bilinears in the
charges\footnote{Note that these bilinears are not positive 
definite as $ O(6,22;\ZZZ)$-invariant matrices have both 
positive and negative eigenvalues.}: $  Q^2, \,P^2 ,\, Q\cdot P $.

Our goal is to compute the index $B_6(Q,P)$. The computation is done 
in the dual frame: type IIB on 
$K3\times  S^1\times \wt S^1$,
where $S^1$ and $\wt S^1$ 
represent two circles which are not factored 
metrically.\footnote{The problem with carrying out 
this computation in heterotic frame is that there 
the system will contain $NS5$-branes, and the 
coupling constant diverges at the core of 
these branes.}  In this
frame, 
we compute $B_6$ for 
a rotating D1-D5-p system\cite{9602065} in  Kaluza-Klein (KK)
monopole (or equivalently Taub-NUT) background. 
More specifically, we take a system containing \cite{0505094}
\begin{enumerate}
\item one KK monopole along $\wt S^1$;
\item one D5-brane wrapped on $K3\times S^1$;
\item $(\wt Q_1+1)$ D1-branes wrapped on $S^1$;
\item $-n$ units of momentum along $S^1$;
\item $J$ units of momentum along $\wt S^1$.
\end{enumerate}
The momentum along $\wt S^1$ appears as an angular
momentum at the center of the Taub-NUT space \cite{0503217}. Thus,
macroscopically, the system describes a 
rotating BMPV black hole \cite{9603078} at the center of 
the Taub-NUT space \cite{0505094}. 
In the weak coupling limit, the dynamics is 
given by that of a
system of decoupled harmonic oscillators, and an exact computation
of $B_6$ is possible.
The result is then expressed in terms of the T-duality invariant
bilinears $ Q^2, \, P^2,\, Q\cdot P $ in the original heterotic frame,
using the fact that the system described above 
has
\be \label{echarges}
Q^2 = 2n, \qquad P^2 = 2\wt Q_1, \qquad Q\cdot P = J\, .
\ee
If $Q^2$, $P^2$ and $Q\cdot P$ were
the only T-duality invariants, \i.e., if any two dyons with the
same $Q^2$, $P^2$ and $Q\cdot P$ had been related to each other
by a T-duality transformation, then the result for $B_6(Q,P)$
for the specific system described above will give the result for
all dyons in the theory. However it turns out that this is not
quite correct. Nevertheless, any charge vector satisfying the
condition\cite{0702150}
\be \label{eaddi}
\gcd\{Q_i P_j - Q_j P_i\,, \quad 1\le i,j\le 28\} = 1\,,
\ee
can be related to the above system by a T-duality
transformation\cite{0712.0043}. Thus the formula
we quote below is valid only for this special class of 
charges.
We shall briefly comment on the other charge vectors in
\S\ref{soth}.

Let us denote by $B_6(\wt Q_1, n, J)$ the sixth helicity
trace associated with
the system described above.
We define the partition function as:
\be \label{pf}
Z(\rho,\sigma,v) = \sum_{\wt Q_1,n,J} \, (-1)^{J}\,
B_6(\wt Q_1,n,J)\, 
e^{2\pi i (\wt Q_1\rho + n\sigma + J v)}\,.
\ee
The computation of $Z$ proceeds as follows. In the weakly coupled
type IIB description, the low energy dynamics of the system
is described by three weakly interacting pieces:
\begin{enumerate}
\item The closed string excitations around the KK monopole.
\item The  dynamics of the D1-D5 center of mass coordinate in the 
KK monopole background.
\item The motion of the D1 branes along $K3$.
\end{enumerate}
The dyon partition function is obtained as the product of the 
partition functions of these three subsystems 
\cite{0605210}.\footnote{A factor of $(-1)^{J+1}$ in \refb{pf} was
missed in \cite{0605210}. The $(-1)^J$ factor arises because in five
dimensions, at the center of the KK monopole, we have $(-1)^F
=(-1)^{J+2h}$ instead of $(-1)^{2h}$\cite{0508174}. An overall
factor of $-1$, which has been absorbed in the definition of $B_6$
in \refb{pf}, arises from the partition function of the quantum
mechanics describing the D1-D5-brane motion in the KK monopole
background\cite{0708.1270}. A detailed derivation of many of
the results given in this section has been reviewed in 
\cite{0708.1270}.}
The analysis can be
simplified by taking the size of $S^1$ 
to be large compared to other dimensions, so that 
we can regard each subsystem
as a 1+1 dimensional CFT. Since BPS condition forces the modes carrying 
positive momentum along $S^1$ (right-moving modes) to be frozen into 
their ground state, only left-moving modes can be excited.
We shall now describe the contribution to $Z$ from each
subsystem.

First consider the fields describing the dynamics of KK 
monopole. These include
\begin{enumerate}
\item 3 left-moving and 3 right-moving bosons arising from its motion 
in the 3 transverse directions;
\item  2 left-moving and 2 right-moving bosons arising from the 
components of 2-form fields along the harmonic 2-form in 
Taub-NUT space \cite{brill,pope};
\item 19 left-moving and 3 right-moving bosons,
arising from the components of the 4-form field along the wedge 
product of the harmonic
2-form on Taub-NUT and a harmonic 2-form on $K3$;
\item 8 right-moving goldstino fermions  
associated with the eight supersymmetries
which are broken by the KK monopole.
\end{enumerate}
Since the right-moving modes are frozen into their ground state,
the contribution to the partition function 
from the KK-monopole 
dynamics, after separating out the contribution from
fermion zero modes which go into the helicity trace, is 
equal to that of 24 left-moving bosons \cite{0605210}:    
\be 
\label{kk}
Z_{KK} = e^{-2\pi i \sigma } \prod_{n=1}^\infty \left\{(1 - 
e^{2\pi i n \sigma })^{-24}\right\}\,.
\ee
The overall factor of $e^{-2\pi i\sigma}$ is a reflection of the fact
that the ground state of the Kaluza-Klein monopole carries a
net momentum of $1$ along $S^1$.

The dynamics of the D1-D5 center of mass motion in the KK monopole
background is described by a supersymmetric sigma model with 
Taub-NUT space as the target space. By taking the size of the 
Taub-NUT space to be large, we can take
the oscillator modes to be those of a free field theory, but the
zero mode dynamics is described by a supersymmetric quantum mechanics
problem. The contribution is found to be \cite{0605210}
\be \label{d1d5}
Z_{CM}= e^{-2\pi i v}
 \prod_{n=1}^\infty \left\{
(1 - e^{2\pi i n \sigma})^4 \, ( 1 - e^{2\pi i n \sigma + 2\pi i
 v})^{-2} \,  ( 1 - e^{2\pi i n \sigma - 2\pi i
 v})^{-2}\right\} \, e^{-2\pi i  v} \,  (1 - e^{-2\pi i  v})^{-2}\,. 
\ee

The third component comprises D1-brane motion along $K3$. This can be 
computed as outlined below \cite{9608096}:  
\begin{enumerate}
\item First consider a single D1-brane, wrapped $k$ times along $S^1$
and carrying  fixed momenta along $S^1$ and $\wt S^1$. The dynamics 
of this system is described by a supersymmetric sigma model with 
target space $K3$. The number of states of this system can be counted 
by the standard method of going to the orbifold limit.
After removing a trivial degeneracy factor associated with fermion zero mode
quantization,
the net number of bosonic minus fermionic states, carrying momentum
$-l$ along $S^1$ and $j$ along $\wt S^1$, is given by $c(4lk-j^2)$,
where $c(n)$ is defined as:
\ben \label{elliptic}
F(\tau,z)  &\equiv& 8 \left[ {\vartheta_2(\tau,z)^2
\over \vartheta_2(\tau,0)^2} +
{\vartheta_3(\tau,z)^2\over \vartheta_3(\tau,0)^2}
+ {\vartheta_4(\tau,z)^2\over \vartheta_4(\tau,0)^2}\right]\,, \\
F(\tau,z) &=&\sum_{j\in \zzz, n }
c(4n-j^2)\,  e^{2\pi i n\tau+2\pi i z j} \,.
\een
Physically, $c(4n-j^2)$ counts the number of BPS states in the
supersymmetric sigma model with target space $K3$ with
$L_0=n$ and $\JJ_3=j/2$, where $\JJ_3$ denotes the third
component of the $SU(2)$ R-symmetry current.  
\item A generic state contains multiple D1-branes of
this type, carrying different amounts of winding along $S^1$
and different momenta along $S^1$ and $\wt S^1$.
The total number of states can be determined from the result of step 
1 by simple combinatorics.
\end{enumerate}
The net contribution to the partition function 
from D1-brane motion along $K3$ is \cite{9608096}:
\be \label{d1}
Z_{D1}=e^{-2\pi i  \rho }
\prod_{l,j,k\in Z\atop
k> 0,l\ge 0}\Bigg\{ 
 1 - e^{2\pi i( l\sigma 
 + k \rho + j v)} \Bigg\}^{-c(4lk -j^2)}\,,
\ee

After taking the product of the component partition 
functions \refb{kk}, \refb{d1d5}     
and \refb{d1}, we get \cite{0605210}
\be \label{fullpf}
Z= e^{-2\pi i  (\rho +\sigma + v)}
\prod_{l,j,k\in Z\atop
k\ge 0,l\ge 0, j<0 \, {\rm for} \,
k=l=0}\Bigg\{ 
 1 - e^{2\pi i( l\sigma 
 + k \rho + j v)} \Bigg\}^{-c(4lk -j^2)}\,,
 \ee
where we have used the explicit values of $c(u)$ to express
the contribution from \refb{kk} and \refb{d1d5}
in terms of
$c(n)$. Indeed these two factors give the $k=0$ term in
\refb{fullpf}. Eq.\refb{fullpf}
can be expressed as
\be 
Z(\rho,\sigma,v) = 1/ \Phi_{10}(\rho,\sigma,v)\,.
\ee
Here $\Phi_{10}$ is a well known function, known
as the weight 10 Igusa cusp form of 
$Sp(2,\ZZZ)$\cite{igusa1,igusa2}.\footnote{$Sp(2,\ZZZ)$ 
includes the $SL(2,\ZZZ)$ S-duality group, but it
is a much bigger group than the S-duality group of 
string theory. Thus it is not completely understood why $Z$ has 
$Sp(2,\ZZZ)$ symmetry (see \cite{0506249,0612011,0808.1746} 
for some attempts in this
direction). In fact, this property of $Z$ comes out 
at the very end after combining the results from the individual 
subsystems. But once we arrive at this final form, these symmetries 
can be conveniently used to analyse the asymptotic behaviour of $Z$.} 
 The formula for $Z$ given 
above was conjectured in \cite{9607026}.

Eq.\refb{pf}
can be inverted to express $B_6(\wt Q_1,n,J)$ as
\be \label{pffourier1}
-B_6(\wt Q_1,n,J) =  (-1)^{J+1}\, \int d\rho d\sigma dv \, 
e^{-2\pi i (\wt Q_1\rho + n\sigma + J v)}\, Z(\rho,\sigma,v) \, .
\ee
We shall express this in a more duality invariant notation 
using \refb{echarges}:
\be \label{edual}
-B_6(Q,P) =  (-1)^{Q\cdot P+1}\, \int d\rho d\sigma dv \, 
e^{-\pi i (P^2\rho + Q^2\sigma + 2Q\cdot P v)}\, Z(\rho,\sigma,v)
\, .
\ee

\subsection{Asymptotic expansion}
In order to compare \refb{edual} with the black 
hole entropy, we need
to find its behaviour for large $Q^2$, $ P^2$, 
$Q.P$. It turns out that this is controlled by the behaviour of
$Z$ at its poles, which in turn are  
at the zeroes of $\Phi_{10}$ \cite{9607026}. 
The location of the zeroes of $\Phi_{10}$ as well as the behaviour
of $\Phi_{10}$ around these zeroes can be determined using its
modular properties.
We perform one 
of the three integrals
using the residue theorem, picking up contributions from various poles.
The leading contribution comes from the pole at \cite{9607026}
\be \label{pole}
(\rho\sigma-v^2) + v = 0 \,.
\ee
After picking up the residue at this pole, we are left with
a two dimensional integral:
\be \label{pffourier2}
-B_6(Q,P) \simeq \int {d^2\tau\over
\tau_2^2} \, e^{F(Q^2,P^2,Q.P,\tau_1, \tau_2)} \,,
\ee
where $(\tau_1, \tau_2)$ parametrize the locus of the zeroes
of $\Phi_{10}$ at \refb{pole} in the $(\rho,\sigma,v)$ space
and 
\be\label{edeff}
F = {\pi\over 2\tau_2} (Q-\tau P)\cdot (Q-\bar\tau P) 
-24\ln \eta(\tau) - 24\ln\eta(-\bar\tau) - 12 \ln (2\tau_2)
+\ln \left[ 
26 + {\pi\over \tau_2} (Q-\tau P)\cdot (Q-\bar\tau P) 
\right]\, .
\ee
We evaluate this integral by the saddle point method. 
We expand $ F$ around its 
extremum and carry out the integral using perturbation theory. 
If we consider a limit in which we scale all the charges by
some large parameter $\Lambda$, then the perturbation expansion
around the saddle point generates a series in inverse power of
$\Lambda^2$, with the leading semi-classical result
being of order $\Lambda^2$.

Applying the above procedure, first of all we 
find that, for large charges, $-B_6(Q,P)$ is
positive \cite{0708.1270} (\i.e., $B_6(Q,P)$
is negative). Furthermore \cite{0412287,0601108}:
\be \label{asym} 
\ln |B_6(Q,P)| = \pi\sqrt{Q^2 P^2 - (Q.P)^2} - \phi\left({Q.P\over P^2}, 
{\sqrt{Q^2 P^2 - (Q.P)^2}\over P^2}\right)
+\OO\left( {1\over Q^2,P^2, Q.P}\right)\,,
\ee
where
\be 
\phi(\tau_1, \tau_2)  \equiv
12\ln \tau_2 +24\ln\eta(\tau_1
+i\tau_2) +
24\ln\eta(-\tau_1+i\tau_2) \,.
\ee
The first term, $\pi\sqrt{Q^2P^2-(Q\cdot P)^2}$, is indeed the
Bekenstein-Hawking entropy of the black hole \cite{9507090,9508094,9512031}. 
The macroscopic
origin of the other terms will be discussed in \S\ref{slook}.

\subsection{Walls of marginal stability}  \label{swall}

Our result for the D1-D5-KK monopole system was derived for 
weakly coupled type IIB string theory. However, as we move 
around in the moduli space, we may hit walls of marginal 
stability, at which the quarter-BPS dyon under consideration 
becomes unstable against decay into a pair of half-BPS dyons. 
At these walls, the index jumps, and hence we cannot trust
our formula on the other side of the wall.
It turns out, however, that with 
the help of S-duality, we can always bring the moduli to a domain 
where the type IIB theory is in the weakly coupled domain and 
we can trust our original formula.
The net outcome of this analysis is that, in different domains, the
index is given by the formula:
\be \label{deg}
-B_6(Q,P) = (-1)^{Q.P+1}\, \int_C d\rho d\sigma dv \, 
e^{-\pi i (P^2\rho + Q^2\sigma + 2Q.P v)} /  \Phi_{10}(\rho,\sigma,v) \,,
\ee
where $C$ denotes the choice of `contour' 
that picks a 3 real dimensional 
subspace
of integration in the 3 complex dimensional space:
\be \label{econt}
Im(\rho)=M_1, \quad Im(\sigma)=M_2, \quad Im(v)=M_3,
\quad 0\le Re(\rho), Re(\sigma), Re(v)\le 1\, .
\ee
The three real numbers $(M_1, M_2, M_3)$, which specify
the choice of the contour 
$C$, depend on the domain in the moduli space
where we compute the index \cite{0702141,0702150,0706.2363}.
For example in the weak coupling     
limit of type IIB string
theory, for the system we have analyzed, we have
$M_1, M_2>>1$, $1<< |M_3|<< M_1, M_2$ and the sign
of $M_3$ is positive or negative depending on whether
the angle between $S^1$ and $\wt S^1$ is larger or
smaller than $\pi/2$ \cite{0605210,0609109}.
The jumps in the index, across the walls of marginal stability, are
encoded in the residues at the
poles in $Z$ that we encounter while deforming the contour
corresponding to one domain to the contour corresponding to the
other domain. There is a precise correspondence between
different walls of marginal stability and different poles of
$Z$. For the decay $(Q,P)\Rightarrow (Q,0) +
(0,P)\,$, the associated wall is at 
$v=0$\cite{0605210,0607155,0609109,0702141,0702150}. This, together
with the S-duality invariance of the theory, tells us that for the
wall associated with the decay
\be \label{edec1}
(Q,P)\Rightarrow (\alpha Q + \beta P, \gamma Q+\delta P)
+ ((1-\alpha) Q - \beta P, -\gamma Q+(1-\delta) P)\, ,
\ee
the corresponding pole is at
\be \label{edecay2}
\gamma\rho -\beta \sigma + (\alpha -\delta) v = 0\, .
\ee
A precise formula giving $(M_1,M_2, M_3)$ in terms of the moduli
and charges can be found in \cite{0706.2363}. We should keep in mind,
however, that the result is independent of $(M_1, M_2, M_3)$ as
long as changing them does not make the contour cross a pole.

On the black hole (macroscopic) side, these
jumps correspond to (dis-)appearance of two-centered black holes as we 
cross walls of marginal stability. There is a precise match between
the $B_6$ index of 2-centered black holes carrying charges given
on the right hand side of \refb{edec1}, and the change
in $B_6(Q,P)$ computed from the residues at the poles 
\refb{edecay2} \cite{0705.3874,0706.2363}.

In this context, we would like to mention that the 
changes in the index across the walls of marginal
stability are subleading,   
as these give corrections which grow as exponentials
of single power of the charges.
This is related to the fact that only decays of a 1/4-BPS dyon into 
half-BPS dyons contribute to the wall crossing in
an $\NN=4$ supersymmetric string 
theory \cite{0707.1563,0803.3857,0903.2481}.
However the contribution from the
multi-centered solutions 
can become significant when we study dyons
in $\NN=2$ supersymmetric string     
theories \cite{0702146}.  

\subsection{Other duality orbits} \label{soth}

We have already 
said that the results given above 
are valid for a subset of dyons satisfying 
the condition \refb{eaddi}. These
can
be related via duality transformation to the D1-D5-p-KK system
analyzed here. But we would like to see if we can say something 
about the dyons which are outside these
duality orbits, \i.e., which have\cite{0702150}
\be \label{eadd3}
\gcd\{Q_i P_j - Q_j P_i\,, \quad 1\le i,j\le 28\} = r\,,
\ee
for some integer $r>1$.
These dyons can be related to a system of IIB 
on $K3\times S^1\times
\wt S^1$ with \cite{0702150, 0712.0043, 0801.0149}
\begin{enumerate}
\item 1 KK monopole along $\wt S^1$,
\item $r$ D5-branes wrapped on $K3\times S^1$,
\item $(\wt Q_1 + 1)\, r$ D1-branes wrapped on $S^1$,
\item $-n$ units of momentum along $ S^1$,
\item $ rJ$ units of momentum along $ \wt S^1$.
\end{enumerate}
If we can compute the $B_6$ index for these dyons, we can use this
to compute the $B_6$ index of any other dyon.
This has not yet been done from first principles, but a guess has
been made by 
requiring that wall crossing is controlled by the residues at the poles
of the partition function as in the $r=1$ case. 
In the domain of the moduli space where 2-centered black holes are
absent, the proposal for the $B_6$ index for these dyons 
is \cite{0802.1556}
\be \label{epropr}
\sum_{s|r} s\, B_6\left(\wt Q_1{r\over s}, n, J{r\over s}\right)\,,
\ee
where $B_6(\wt Q_1,n,J)$ is the function 
defined in \refb{pffourier1}. 
An effective string model for arriving at this
result has been suggested in \cite{0803.2692}, 
but this has not been derived completely  
from first principles.
Note that for large charges, the contribution from the
$s>1$ terms grow as  
$\exp(\pi\sqrt{Q^2 P^2 - (Q\cdot P)^2}/s)$
and hence are exponentially suppressed compared
to the leading $s=1$ term. Thus the result
for the index reduces to that for the $r=1$ case up to exponentially
suppressed corrections.

\subsection{Generalization I: Twisted index}  \label{sg1}

Let us take type IIB theory on $K3\times S^1\times \wt S^1$. On
special 
subspaces of the moduli space of $K3$, we encounter enhanced discrete 
symmetries which preserve the holomorphic
(2,0)-form on $K3$ \cite{9508144,9508154}. 
Thus these symmetries 
commute with supersymmetry.
Let us work on such a subspace of the moduli space with
a $\ZZZ_N$ discrete symmetry generated by $g$.
In this subspace, we can define a twisted index:
\be \label{twistindex} 
B^g_6 = {1\over 6!}
Tr \lbrace (-1)^F (2h)^6 g\rbrace\,.
\ee
This can be calculated using the same 
method described earlier
by keeping track of the $g$ quantum numbers of the various modes contributing to the partition function.
The final result takes the form \cite{0911.1563}:
\be \label{b6}
B^g_6(Q,P) = (-1)^{Q.P}\, \int_C d\rho d\sigma dv \, 
e^{-\pi i (P^2\rho + Q^2\sigma + 2Q.P v)}\, Z^g(\rho,\sigma,v) \,,
\ee
where the functions $Z^g$ are known explicitly.
They also turn out to have nice 
modular properties and
poles in the complex $(\rho,\sigma,v)$ space.\footnote{General
discussion on such modular forms can be found in 
\cite{borcherds,9504006,ibu1,ibu2,ibu3,eichler,skor,rama}.}
As a result, we can find the behaviour of this index for large 
charges by the same method
described earlier. The important difference is that now there are
no poles at \refb{edecay2}. Instead the poles are at \cite{0911.1563}
\ben \label{epoleg}
&& n_2 (\rho\sigma-v^2) - m_1 \rho+n_1\sigma +m_2 + jv=0\,, \qquad
m_1 n_1 + m_2 n_2 + {1\over 4} j^2 = {1\over 4}\,, \nonumber \\
&& \qquad m_1, n_1, m_2\in\ZZZ \,, \qquad
j\in 2\ZZZ+1 \,, \qquad n_2\in N\ZZZ\, .
\een
The leading contribution now comes from the poles at
\refb{epoleg} with $n_2=N$,
and the answer in the large charge limit is \cite{0911.1563}:
\be
\ln |B_g^6(Q,P)| = \pi 
\sqrt{Q^2 P^2 - (Q.P)^2} / N  +\OO(1)\,.  
\ee
A macroscopic explanation of this result will be given in
\S\ref{stw}.

\subsection{Generalization II: CHL models}
\label{sg2}  

We again start with type IIB string theory on $ K3\times S^1
\times \wt S^1$ with a $\ZZZ_M$ symmetry generated by
$\wt g$ as described in
\S\ref{sg1}, but this time we take an orbifold of this theory by $\wt g$ 
accompanied by 
$2\pi/M$ shift along $S^1$.\footnote{The $\ZZZ_M$ symmetries
are chosen from the same set as the $\ZZZ_N$ symmetries of the
\S\ref{sg1}, but we are using a different label since in the
next section we shall combine the analysis of \S\ref{sg1} and this
subsection.}
This generates a new
class of $\NN=4$ supersymmetric string theories
known as CHL models \cite{9505054,9506048}. The orbifold operation
removes some of the $U(1)$ gauge fields. Thus, in general, CHL
models have
$(r+6)$ U(1) gauge fields with $r < 22$,     
and $Q$ and $P$ are $(r+6)$ dimensional 
vectors.\footnote{Since 6 of the U(1) gauge fields
represent graviphoton fields, they must exist in all $\NN=4$
supersymmetric string theories.}
The
precise value of $r$ depends on $M$, -- the order of the
orbifold group.
The T-duality group is a discrete subgroup of $O(6,r)$ with 
$Q$ and $P$ transforming as vectors of $O(6,r)$. Thus $O(6,r)$
invariant bilinears $Q^2$, $P^2$ and $Q\cdot P$ are T-duality
invariants. 

In this theory we can take the same D1-D5-KK monopole system
as considered earlier since all of these configurations, as well
as momenta along $S^1$ and $\wt S^1$, are invariant under the
orbifold group.
The index $ B_6$ in this theory can be calculated in the 
same way as before, keeping track
of the $\wt g$  quantum numbers of the various modes, and the effect
of the orbifold projection. The result of this computation 
is \cite{0605210}:
\be \label{chlb6}
B_6(Q,P) = (-1)^{Q.P}\, \int_C d\rho d\sigma dv \, 
e^{-\pi i (P^2\rho + Q^2\sigma + 2Q.P v)}\, \wt Z^g(\rho,\sigma,v) \,,
\ee
where $\wt Z^g(\rho,\sigma,v) $ is yet another new function, also with
nice modular properties and poles in the $(\rho,\sigma,v)$ space.
We find that its behaviour for large charges is given by:
\be \label{echlfor}
\ln |B_6(Q,P)| = \pi\sqrt{Q^2 P^2 - (Q.P)^2}  -   
\phi\left({Q.P\over P^2}, \sqrt{Q^2 P^2 - (Q.P)^2\over P^2}\right)
+\OO\left( {1\over Q^2,P^2, Q.P}\right)\,,
\ee
where
\be \label{eend}
\phi(\tau_1, \tau_2) \equiv
(k+2)\ln \tau_2 + \ln~g(\tau_1
+i\tau_2) +
 \ln~g(-\tau_1+i\tau_2) \,.
\ee
Here $k$ are known numbers and $ g(\tau)$ are known functions,
depending on the choice of $M$.
This generalizes \refb{asym}.
Furthermore, in each case we have $B_6(Q,P)<0$.
The macroscopic origin of \refb{echlfor} will be explained
in \S\ref{slook}, and the macroscopic explanation of the sign
of $B_6$ will be given 
in \S\ref{sdeg}.

Note that unlike in the case of heterotic string theory 
on $T^6$, in this case the duality orbits have not been
completely classified. As a result, two vectors with the
same values of $Q^2$, $P^2$ and $Q\cdot P$ are not
necessarily related by a duality transformation. Our result
for the index, given in \refb{chlb6}, holds only for those
charge vectors which can be related by a duality  
transformation to the specific
D1-D5-KK monopole system for which we have carried out
our analysis.

\subsection{Generalization III: Twisted index in CHL models}
\label{sg3}

Next we consider a special subspace of the moduli space on which  
type IIB string theory on $K3\times S^1\times \wt S^1$ has
a $\ZZZ_M\times \ZZZ_N$ discrete symmetry that commutes
with supersymmetry.
Let $\wt g$ and $g$ be the generators of $\ZZZ_M$  
and $\ZZZ_N$ respectively.
Let us now take an orbifold of this theory by a $\ZZZ_M$ symmetry
generated by $\wt g$ together with $ 1/M$ unit of shift along
$ S^1$. Here $g$ still generates a symmetry of the theory.
We now define:
\be
B^g_6 = {1\over 6!}
\, Tr \lbrace (-1)^F (2h)^6 g\rbrace\,. 
\ee
The computation of the above index gives the result \cite{1002.3857}
\be
B^g_6(Q,P) = (-1)^{Q.P}\, \int_C d\rho d\sigma dv \, 
e^{-\pi i (P^2\rho + Q^2\sigma + 2Q.P v)}\, \wh 
Z^{g,\wt g}(\rho,\sigma,v)\,,     
\ee
where $\wh Z^{g, \wt g}$ is yet another set of     
functions, also with nice modular properties and
poles in the complex $(\rho,\sigma,v)$ space.
Its behaviour for large 
charges is found to be
\be
\ln |B^g_6(Q,P)| = \pi 
\sqrt{Q^2 P^2 - (Q.P)^2} / N +\OO(1)\,.
\ee
A macroscopic explanation of this result will be given in
\S\ref{stw}.

\subsection{Generalization IV: Twisted index in type II 
string compactification}

The analysis described above has also been generalized to untwisted
and twisted
indices in type II string compactifications
on $T^6$ and its asymmetric orbifolds. 
We shall not describe the
analysis here; they can be found in \cite{0607155,0609109,0911.1563,1002.3857}. 
The general feature
of all these models is that a $\ZZZ_N$ twisted index $B^g_6$ grows as
\be \label{egen1}
\ln |B^g_6(Q,P)| = \pi 
\sqrt{Q^2 P^2 - (Q.P)^2} / N + \OO(1)\, .
\ee
This includes the case of $N=1$, \i.e., the untwisted index, for which
$\ln |B_6(Q,P)| \simeq \pi 
\sqrt{Q^2 P^2 - (Q.P)^2}$.
Macroscopic explanation for these results is the same as that for
the black holes in heterotic string theories, and hence we shall not
discuss these cases separately.

\subsection{Other systems} \label{sothersystem}

Finally we must mention that besides the systems described above, there
are other systems for which the microscopic results are known exactly.
These include the following:
\begin{enumerate}
\item
A special mention must be given to the untwisted index in type II
string theory on $T^6$. This theory has $\NN=8$ supersymmetry and
the black holes with finite area event horizon are 1/8-BPS.
Thus the relevant helicity trace index is $B_{14}$. 
For a class of 1/8 BPS states in this theory 
the microscopic result for the index $B_{14}$ is known 
exactly\cite{0506151,0508174,0803.1014,0804.0651}.  
In this case the theory has 12 NSNS sector gauge fields
and 16 RR sector gauge fields. If we consider a state carrying only NSNS
sector electric and magnetic charges $Q$ and $P$ and satisfying the condition
\refb{eaddi} with $1\le i,j\le 12$,
then the result for $B_{14}$ is:
\be \label{eb14}
B_{14} = (-1)^{Q\cdot P}\,
\sum_{s|Q^2/2, P^2/2, Q\cdot P} s\, \wh c(\Delta / s^2) 
\ee
where 
\be \label{eb14a}
\Delta = Q^2 P^2 - (Q\cdot P)^2\, ,
\ee
and $\wh c(n)$ is defined via the expansion
\be \label{eb14b}
-\vartheta_1(z|\tau)^2 \, \eta(\tau)^{-6} \equiv \sum_{k,l} \wh c(4k-l^2)\, 
e^{2\pi i (k\tau+l z)}\, .
\ee
$\vt_1(z|\tau)$ and $\eta(\tau)$ are respectively 
the first Jacobi theta function and Dedekind function. Given this result
we can derive the result for $B_{14}$ for many other states using the
U-duality symmetries of the theory, but they do not span all possible
charge vectors in the theory\cite{0804.0651}. 
For large charges one finds that $B_{14}$
is negative and\cite{0908.0039}
\be \label{eb14c}
\ln |B_{14}| = \pi\sqrt{\Delta} - 2\ln \Delta\, .
\ee
The first term on the right hand side is the Bekenstein-Hawking
entropy of a black hole carrying the same charges. The origin of the
logarithmic
correction on the macroscopic side will be discussed in \S\ref{slog}.
\item For a class of five dimensional theories, including type II string theory
compactified on $T^5$ or $K3\times S^1$ and their orbifolds preserving
sixteen supersymetries, the microscopic results for the index is known. These
systems are in fact closely related to the four dimensional systems discussed
above since the latter are constructed by placing the five dimensional system
at the center of a Taub-NUT space. 
We shall discuss the case of CHL orbifolds of 
type IIB  on $K3\times S^1$ preserving 16 supersymmetries,
but similar results are also available for type IIB on $T^5$. In
this case a general rotating black hole carries two angular momenta
$J_{3L}$ and $J_{3R}$ labelling the Cartan generators of the rotation
group $SO(4)=SU(2)_L\times SU(2)_R$. However supersymmetry requires
one of the angular momenta (which we shall take to be $J_{3R}$ to
vanish). The microstates of the black hole at weak coupling however do
not necessarily have vanishing $J_{3R}$, 
and the relevant protected index that counts these
microstates is given by
\be \label{edefindalt}
\wt d_{micro}(n,Q_1,Q_5,J)\equiv
-{1\over 2!}\,
\wt{Tr}\left[ (-1)^{2 J_{3R}} \, (2 J_{3R})^2
\right]\, ,
\ee
where $Q_1$, $Q_5$ and $n$ denote respectively the charges corresponding
to D1-brane wrapping along $S^1$, D5-brane wrapping along $K3\times S^1$,
and momentum along $S^1$, and the trace is  taken over states 
carrying fixed $Q_1$, 
$Q_5$, $n$
and $J_{3L}=J/2$,  but
different values of $\vec J_L^2$, $J_{3R}$ and $\vec J_R^2$.
One can also consider another protected index 
$d_{micro}(n,Q_1,Q_5,J)$ where $\vec J_L^2$
is also fixed at $J/2(J/2+1)$. There two indices are related by the simple
formula
\be \label{e5d1}
d_{micro}(n,Q_1,Q_5,J) = \wt d_{micro}(n,Q_1,Q_5,J) 
- \wt d_{micro}(n,Q_1,Q_5,J+2)\, .
\ee
When $Q_1$ and $Q_5$ are relatively prime,
the result for $\wt d_{micro}$ for type IIB on $K3\times S^1/\ZZZ_N$ is
given by\cite{1109.3706} 
(see \cite{9608096,0807.0237,0807.1314,1009.3226} for the $N=1$ case)
\be \label{e5d2}
\wt d_{micro}(n,Q_1,Q_5,J) = (-1)^{J}\, \int_C d\rho d\sigma dv \, 
e^{-2\pi i (Q_1Q_5\rho + n\sigma + Q.P v)}\, \check Z^g(\rho,\sigma,v) \,,
\ee
where $\check Z^g$ is a function that is closely related to the function
$Z^g$ that appears in \refb{chlb6}.
Using \refb{e5d1} and \refb{e5d2} one can calculate the asymptotic
behaviour of $\wt d_{micro}$ and $d_{micro}$ in the limit when
the charges and angular momenta are large. It turns out that
besides the leading contribution which agrees with the Bekenstein-Hawking
entropy of the corresponding black holes, there are linear and
logarithmic corrections
to the entropy. The linear corrections arise from a shift in the definition
of the charge and was shown to agree between the microscopic and
the macroscopic side in \cite{0807.0237}.
The logarithmic corrections will be discussed in table \ref{t2} in \S\ref{slog}.

\end{enumerate}

\sectiono{Macroscopic analysis}
\label{macroscopic}

Our next goal is to 
\begin{itemize}
\item develop tools for computing the entropy of extremal black holes
including stringy and quantum corrections,
\item relate this entropy to the helicity trace index,  
\item apply it to  black holes carrying the same charges for which we
have computed the microscopic index, and
\item compare the macroscopic results with the microscopic results.
\end{itemize}
In this section, we shall mainly address the 
first and the second issues, \i.e.,  
find a general formula for computing the
black hole degeneracy and the helicity trace
on the macroscopic side. 
Some aspects of the third and the fourth  
issues will be discussed
in \S\ref{slook}, but we postpone the major part of this to
\S\ref{sapp}.
Since $AdS_2$ space will play a crucial role in our
analysis, we begin by describing some aspects of $AdS_2$
space.

\subsection{What is $AdS_2$ ?}

Take a three dimensional space labelled by coordinates
$(x,y,z)$ and metric
\be \label{metricflat} 
ds^2 = dx^2 - dy^2 - dz^2\,.
\ee
 $AdS_2$ may be regarded as a two dimensional
 Lorentzian space embedded in this 3-dimensional  space
via the relation:
\be 
x^2 - y^2 - z^2 = -a^2\,,
\ee
where $a$ is some constant giving the radius of $AdS_2$.
Clearly, this space has an SO(2,1) isometry.
 
Introducing the independent coordinates $(\eta,t)$ such that
 \be 
 x = a \, \sinh \eta \cosh t, \quad y = a \, \cosh \eta, \quad
z = a \, \sinh \eta \sinh t\,,
\ee
we can write
\be 
dx^2 - dy^2 - dz^2 = a^2 (d\eta^2 - \sinh^2 \eta \, dt^2)\,.
\ee
Finally, defining 
\be 
r=\cosh\eta\,,
\ee
the metric for $AdS_2$ can be expressed as:
\be \label{metricads}
ds^2 = a^2\left[{dr^2\over r^2-1} - (r^2-1) dt^2
\right], \quad r\ge 1 \,.
\ee
Using a change of coordinates, one can show that the apparent singularity
at $r=1$ is a coordinate singularity, and one can continue the
space-time beyond $r=1$ to generate what is known as global $AdS_2$
space-time. This will not play any direct role    
in our subsequent discussion.

\subsection{Why $AdS_2$ ?}

The reason that $AdS_2$ plays an important role for extremal
black holes is that all known black holes develop an $AdS_2$ factor in their
near horizon geometry in the extremal limit. In particular, 
the time translation 
symmetry gets enhanced to the $SO(2,1)$ isometry of $AdS_2$.
We shall illustrate how this happens by considering the example of
Reissner-Nordstrom solution in 
$ D=4$. This is described by the metric
 \be
ds^2 =- (1 - \rho_+/\rho) (1 - \rho_-/\rho) d\tau^2 
 + {d\rho^2\over
 (1 - \rho_+/\rho) (1 - \rho_-/\rho)}
 + \rho^2 (d\psi^2 + 
 \sin^2\psi d\phi^2) \,. 
 \ee
Here $\rho_\pm$ are parameters determined in terms of the mass
and charges carried by the black hole. In the extremal limit, $\rho_-\to
\rho_+$. In order to take this limit, we
define:
\be 
 2\lambda=\rho_+-\rho_-, \quad
  t = {\lambda\, \tau \over \rho_+^2}, 
 \quad r={2\rho-\rho_+
 -\rho_-\over 2\lambda}\,,
 \ee
and take $\lambda\to 0$ limit keeping $r,t$ fixed. In this limit, 
the metric takes the form \cite{9707015,9709064,9812073}:
\be 
ds^2 =  \rho_+^2 \left[-(r^2-1) dt^2 + 
   {dr^2\over r^2-1} \right]+
 \rho_+^2 (d\psi^2 + \sin^2\psi d\phi^2)\,.
 \ee
 This describes the space $AdS_2 \times S^2$. One can also verify
 that, in this limit, the near horizon electric and magnetic fields
 are invariant under the isometries of $AdS_2\times S^2$.

We will now postulate that \textit{any extremal black hole
has an $AdS_2$ factor / $ SO(2,1)$ isometry
in the near horizon geometry.} 
This postulate has been partially proved in \cite{0705.4214, 
0803.2998}.
The full near horizon geometry takes the form $AdS_2 \times K$, 
where $K$ is some compact space. $K$ includes not only the 
compact space on which string theory
is compactified (to get a four dimensional theory), but 
also the angular coordinates
({\it e.g.} the $S^2$ factor for spherically symmetric black holes in
four dimensions).

\subsection{Higher derivative corrections}

In string theory, we expect the Bekenstein-Hawking
formula for the black hole entropy to receive
\begin{itemize}
\item
higher derivative corrections arising in classical string theory,   
and
\item 
quantum corrections.
\end{itemize}
Of these, the higher derivative corrections are captured by
Wald's general formula for black hole entropy in any
general coordinate invariant classical theory of 
gravity \cite{9307038,9312023,9403028,9502009}.
Furthermore, this
formula
takes a very simple prescription for
black holes with an $AdS_2$ factor in the near horizon 
geometry \cite{0506177,0508042,0606244}.
We shall illustrate this in the context of spherically symmetric
black holes in four dimensional theories.
In this case, the near horizon geometry has an $AdS_2\times S^2$ factor.
Consider an arbitrary general coordinate invariant theory of gravity
coupled to a set of gauge fields $A_\mu^{(i)}$ and neutral
scalar fields $\{\phi_s\}$.
The most general form of the
near horizon geometry of an extremal black hole, consistent with
the symmetry of $AdS_2\times S^2$, is:
\ben \label{enear}
ds^2 &\equiv & g_{\mu\nu}dx^\mu dx^\nu
= v_1\left(-(r^2-1) dt^2+{dr^2\over r^2-1}\right)   
+ v_2 \left(d\psi^2
+\sin^2\psi d\phi^2\right)\,,\nonumber \\
&& \phi_s =u_s\,, \qquad   F^{(i)}_{rt} = e_i\,,  \qquad
F^{(i)}_{\psi\phi} = {p_i \over 4\pi} \,
\sin\psi\,,
\een
where $v_1$, $v_2$, $\{u_s\}$, $\{e_i\}$ and $\{p_i\}$ are constants.
For this background, the components of the Riemann tensor are given by:
\ben
R_{\alpha \beta \gamma \delta} &=& - v_1 (g_{\alpha\gamma}\,g_{\beta\delta} 
- g_{\alpha\delta} \,g_{\beta\gamma})\quad \mbox{(where $\alpha,\beta,
\gamma, \delta=r,t$)}\,,\nonumber \\
R_{mnpq} &=& v_2 (g_{mp}\, g_{nq} - g_{mq}\, g_{np}) \quad \mbox{(where 
$m,n,p,q=\psi,\phi$)}\,.
\een
The covariant derivatives of
the Riemann tensor,  scalar fields and gauge field strengths
vanish.

Let $\sqrt{-\det g} \, \LL $ be the Lagrangian density
evaluated in the background \refb{enear}.     
We define the functions:
\be \label{entropyfn}
f(\vec u, \vec v, \vec e, \vec p)\equiv
\int d\psi \, d\phi\, \sqrt{-\det g}\,\LL
\,, \qquad \EE (\vec u, \vec v, \vec e, \vec q, \vec p) \equiv 2\,
\pi (\, e_i \, q_i - f(\vec u, \vec v, \vec e, \vec p)\,)\,.
\ee
Then for an extremal black hole of electric
charge $ \vec q$ and magnetic charge
$\vec p$, one finds that \cite{0506177}
\begin{enumerate}
\item the values of $\{u_s\}$,  $ \{e_i\}$,
$ v_1$ and $ v_2$
are obtained by extremizing $\EE (\vec u, \vec v, 
\vec e, \vec q, \vec p)$ with
respect to these variables:
\be \label{eom}
{\p \EE  \over \p u_s}=0\,, \quad {\p
\EE  \over \p v_1}=0\,, \quad  {\p
\EE  \over \p v_2}=0\,, \quad {\p \EE \over \p e_i}=0 \,;
\ee
\item the Wald entropy of the black hole is given by
\be \label{ewald}
S_{BH}=\EE  \, ,
\ee
at the extremum.
\end{enumerate}
Eqs.\refb{eom} follows from the equations of motion and the definition
of the electric charge, while \refb{ewald} follows from Wald's formula
for the black hole entropy.  

These results provide us with \cite{0506177,0508042,0606244}
\begin{enumerate}
\item an algebraic method for computing the entropy of extremal
black holes without solving any differential equation;
\item a proof of the attractor
mechanism \cite{9508072,9602111,9602136,0507096}, \i.e.,
the black hole entropy
is independent of the asymptotic moduli.
\end{enumerate}
However, this approach
does not prove the existence of an extremal black hole
carrying a given set of charges; it works assuming that the solution
exists. 

\subsection{Quantum corrections: A first look} \label{slook}

Next we must address the effect of 
quantum corrections on the black hole entropy.
The first guess would be that we should apply Wald's formula again, 
but replacing the
classical action by the one particle irreducible (1PI) 
action. This will again give a 
simple algebraic method for computing
the entropy once we compute the 1PI action.
However, this prescription is not complete 
since the 1PI action typically has non-local
contribution due to massless states propagating in the loops. In contrast,
Wald's formula is valid for theories with local Lagrangian density.
This is apparent in \refb{entropyfn} where the definition of the
function $f$ requires explicit knowledge of the local Lagrangian
density $\LL$.

Nevertheless, this procedure
has been used to compute corrections to black hole
entropy from local terms in the 1PI action with 
significant success \cite{9711053,9801081,9812082,9904005,9906094,
9910179,0007195,0009234,0012232,0405146}.
If we consider the CHL models obtained by $\ZZZ_N$ orbifold of
type IIB on $K3\times S^1\times \wt S^1$, then at tree level there are 
no corrections to the black hole entropy from the four derivative terms
in the effective action.
But at one loop, 
these theories get corrections proportional
to the Gauss-Bonnet term in the 1PI action\cite{9610237,9708062}:
\be 
\sqrt{-\det g}\, \, \Delta\LL =  -{1\over 64\pi^2}\,
\phi(\tau,\bar\tau) \, \sqrt{-\det g}
\left\{ R_{\mu\nu\rho\sigma}\, R^{\mu\nu\rho\sigma}
- 4 R_{\mu\nu}\, R^{\mu\nu}
+ R^2 \right\}\,,
\ee
where $\tau$ is the modulus of the torus $(S^1\times \wt S^1)$ and
$\phi$ is the same function that appeared in \refb{eend}.
Adding this correction to the supergravity action, we find that the Wald entropy of a black hole in the CHL model is given by \cite{9709064}
\be \label{egauss}
S_{BH} = \pi\sqrt{Q^2 P^2 - (Q.P)^2} - \phi\left({Q.P\over P^2}, 
\sqrt{Q^2 P^2 - (Q.P)^2\over P^2}\right)
+\OO\left( {1\over Q^2,P^2, Q.P}\right)\,, 
\ee
in exact agreement with the 
result \refb{echlfor}  for 
$\ln|B_6(Q,P)|$ for large charges.\footnote{In fact the 
original computation
involved a more refined
version of the 1PI action, where the complete supersymmetric
completion of  the curvature squared terms in the 1PI
action was included in the 
computation \cite{9801081,9812082,9904005,9906094,
9910179,0007195,0009234,0012232,0405146,
0603149,0612225,0808.2627}. 
Surprisingly, the result is the
same as in \refb{egauss}.
Nevertheless, there can be additional four derivative
corrections to the action which could give additional
contribution to the entropy to this order. One expects 
that a suitable non-renormalization theorem will make
these additional contributions vanish, but this has not been
proven so far.}

\subsection{Quantum corrections to horizon
degeneracy} \label{sqef}

Let us denote by $d_{hor}$ the degeneracy associated with the
horizon of an extremal black hole.
We shall now turn to the full quantum computation of $d_{hor}$
from the macroscopic side, and describe a proposal for computing 
quantum corrected
entropy in terms of a path integral of string theory in this near
horizon geometry \cite{0809.3304,0903.1477}.
The steps for computing $ d_{hor}$ are as follows:
\begin{enumerate}
\item Go to the Euclidean formalism by the replacement
$t\to -i\theta$ and represent the 
$AdS_2$ factor
by the metric:
\be \label{eclads}
 ds^2 = v\left((r^2-1)\, d\theta^2+{dr^2\over
r^2-1}\right)\,, \quad 1\le r < \infty \,, \quad \theta\equiv\theta+2\pi\,.
\ee
With the change of variable $r=\cosh\eta=(1+\rho^2)/(1-\rho^2)$, 
we get the metric on a unit disk:
\be 
 ds^2 = v\left(\sinh^2\eta \,\, d\theta^2+ d\eta^2\right)
= {4v\over (1-\rho^2)^2} (d\rho^2 + \rho^2 d\theta^2),
\qquad 0\le\eta<\infty, \quad 0\le\rho< 1\,.
\ee
\item Regularize the infinite volume of $AdS_2$ by putting a cut-off 
$r\le r_0f(\theta)\,,$ for some smooth
periodic function $f(\theta)$.
This makes the $AdS_2$ boundary have a finite length $ L$.
\item Define:
\be \label{qefdfn} 
Z_{AdS_2} \equiv \left\langle \exp[-
i  q_k\ointop d\theta \, A^{(k)}_\theta]
\right\rangle \,,
\ee
where the symbol $\langle ~\rangle $ denotes the unnormalized
path integral over string fields in the
Euclidean near horizon background geometry weighted
by $ \exp[-{\rm \small Action}]$. 
Here $\{q_k\}$ stands for the electric 
charges carried by the black hole, 
representing the
electric fluxes of the U(1)
gauge fields $A^{(k)}$'s through $ AdS_2$. The integral
$\ointop$ runs over the boundary of the infrared regulated
$AdS_2$.

Note that near the boundary of 
$ AdS_2$, the 
$\theta$-independent
solution to the Maxwell's equations has the form:
\be 
A_r=0, \quad A_\theta = C_1 + C_2 r\,,
\ee
where $ C_1$ (chemical potential) represents a normalizable mode and
$ C_2$ (electric charge) represents a non-normalizable 
mode. Hence the path integral 
\refb{qefdfn} must be carried out 
keeping $C_2$ (charge)
fixed and integrating over $C_1$ 
(chemical potential).\footnote{This is different from 
the standard rules in higher dimensional
space-time where the asymptotic value of the gauge field is held
fixed.} 
Another way to motivate this is the following: 
in $AdS_2$, if we try to add charge/mass, it will 
destroy the asymptotic boundary conditions as it is a 
two dimensional spacetime. 
With this new rule, the first order variation of the action will contain
a boundary term besides the terms proportional to the equations
of motion. This boundary term must be cancelled by some other term
in order to have a well-defined path integral. 
The boundary term
$\exp[ -i  q_k\ointop d\theta \, A^{(k)}_\theta]$
precisely serves this purpose.

\item 
Now, by $AdS_2/CFT_1$ correspondence, string theory on $AdS_2\times K$
must be dual to a one dimensional conformal field theory, which we shall call
$CFT_1$, living on the boundary of $AdS_2$.
Furthermore, we must have\footnote{We emphasize 
that here, since the 
boundary theory is on a circle, its partition function  
can be given a 
Hilbert space interpretation. This is not possible in higher
dimensional $AdS_{d+1}$ spaces where the boundary theory lives
on $S^d$.}  
\be \label{eadscft}
Z_{AdS_2} = Z_{CFT_1} = Tr(e^{-LH})\,, 
\ee
where $H$ is the Hamltonian of $CFT_1$ and $L$ is the length
of the boundary circle of the infrared regulated $AdS_2$.
The standard rule of $AdS/CFT$ correspondence also gives us some
insight into how to identify the $CFT_1$, -- it must be given by the infrared
limit of the quantum mechanics that describes the black hole
microstates. Now in all known examples, including the ones discussed
in \S\ref{stwo}, the quantum mechanics describing the dynamics
of the microscopic
system has a finite gap that separates the 
ground states from the  
first excited state.\footnote{Even though the dynamics was described
by a two dimensional CFT, the CFT was compactified on a circle of
finite size, and hence had a gap in its spectrum.} Thus in the infrared
limit $(L\to\infty)$, only the ground states of this quantum mechanics
(in a fixed charge sector)
survive, and $CFT_1$ will consist of a finite number $d_0$ of degenerate
ground states of some energy $E_0$. This gives, from \refb{eadscft},
\be
Z_{AdS_2}  = d_0\, e^{-L\, E_0} \,.
\ee
This suggests that we define $d_{hor}$  to be the finite part of
$Z_{AdS_2}$, defined by expressing $Z_{AdS_2}$ as
\be \label{dhor}
Z_{AdS_2} = e^{C L + O(L^{-1})} \times d_{hor}\,,
\quad \hbox{as $L\to
\infty$} \,.
\ee
Here $C$ is a constant. The above definition of $d_{hor}$ will
be called the 
\textit{quantum entropy function}.
\item Finally we note that, since the $AdS_2$ path integral is
evaluated by keeping fixed the asymptotic value of the electric field
(and hence the electric charge for a given action),
the 
$AdS_2$ path integral 
computes the entropy in the
microcanonical ensemble where all the charges are fixed.
\end{enumerate}

One of the consistency tests this proposal must satisfy is that,
in the classical limit, it should reproduce the exponential of
the Wald entropy.
This can be seen as follows: In
the classical limit,
\ben   Z_{AdS_2} &=& 
 \exp[-{\small \rm Action} -    
i  q_k\ointop d\theta \, A^{(k)}_\theta]
\bigg|_{classical}\nonumber \\
&=& \exp\left[\int dr d\theta \,  \left\{\sqrt{\det g_{AdS_2}} 
\LL_{AdS_2} -i q_k 
F^{(k)}_{r\theta}\right\}\right]\, ,
\een
where $g_{AdS_2}$ is the  metric on $AdS_2$,
and $\LL_{AdS_2}$ is the two dimensional Lagrangian density
obtained after dimensional reduction on $K$ and is evaluated on the
near horizon geometry.\footnote{Note that the Euclidean action
is given by $-\int dr d\theta \sqrt{g_{AdS_2}} \, \LL\,,$ where $\LL$
is the analytic continuation of the Lagrangian density for
Lorentzian signature.} 
Taking the infrared cut-off to be
$\eta\le \eta_0$ for simplicity, using the Euclidean version of the
near horizon background given in \refb{enear}, 
and evaluating the $r,\theta$ integral,
we get,
\ben 
Z_{AdS_2}
&=&\exp\left[ -2\pi \left(q_i e_i 
- \sqrt{\det g_{AdS_2}} \, {\cal L}_{AdS_2}\right) 
(\cosh\eta_0-1)\right] \nonumber \\
&=& \exp\left[2\pi \left(q_i e_i 
- \sqrt{\det g_{AdS_2}} \, {\cal L}_{AdS_2}\right) + C L
+ \OO(L^{-1}) \right]  
\nonumber \\
&=& \exp\left[S_{wald} + C L + \OO(L^{-1})\right]\,,
\een
where 
\be
L=\sqrt v\, \sinh\eta_0\, \Rightarrow  \cosh\eta_0 = L/
\sqrt v + \OO(L^{-1})\,.
\ee
The constant $C$ can receive additional corrections from boundary terms
in the action which we have ignored.
The important point is that these boundary terms do not affect the value
of the finite part, and hence $d_{hor}$ is well defined.

This establishes that $d_{hor}=\exp[S_{wald}]$ in the 
classical limit.

We conclude this section with two comments:
\begin{itemize}
\item By choosing the boundary terms appropriately, we could cancel
the constant $C$ and reinterprete the full partition function $Z_{AdS_2}$
as $d_{hor}$. In the dual $CFT_1$, this corresponds to shifting the ground
state energy by adding appropriate counterterms.
\item
Our interpretation of the $AdS_2$ partition function as the
degeneracy associated with the horizon is based on representing
Euclidean $AdS_2$ as a disk with a single boundary. If instead we
represent it as a strip with two boundaries, with the help
of the standard conformal transformation 
$\tan{w\over 2} ={z-1\over z+1} $, mapping the unit disk 
in the $z=\rho\, e^{i\theta}$ plane to
a strip in the $w$ plane, 
then we have two copies of $CFT_1$ living on the two boundaries
of the strip, each with degeneracy $d_{hor}$. Standard 
argument \cite{0106112}
shows that the Hartle-Hawking
state of this system will represent the maximally entangled
state between these 
two copies of the $CFT_1$, and as a result, $d_{hor}$
can be reinterpreted as the entanglement entropy between the two
boundaries in this state.
This has been verified explicitly in \cite{0710.2956} in the classical limit. 
\end{itemize}

\subsection{Hair contribution} \label{shair}

In general, 
the macroscopic degeneracy, denoted by
$d_{macro}$, can have two kinds of
contributions \cite{0901.0359, 0907.0593}:
\begin{enumerate}
\item From the the degrees of freedom living on the horizon.
\item From the degrees of freedom living outside the horizon 
(hair) \cite{0901.0359,0907.0593}.\footnote{For  
example, 
the fermion zero modes 
associated with the broken supersymmetry generators are always 
part of the hair modes, since the effect of supersymmetry-breaking
by the classical black hole solution can be felt outside the
horizon of the black hole.}
\end{enumerate}
Denoting the degeneracy associated with the horizon 
degrees of freedom  by $d_{hor}$ and those associated with the hair
degrees of freedom by $d_{hair}$, 
we can write down a general 
formula for $d_{macro}$: 
\be \label{emacro}
d_{macro}(\vec Q)=\sum_n\,
\sum_{\{\vec Q_i\}, \vec Q_{hair}\atop \sum_{i=1}^n 
\vec Q_i+ \vec Q_{hair}=\vec Q} 
\, \left\{\prod_{i=1}^n \, d_{hor}(\vec Q_i)\right\}  \,
d_{hair}(\vec Q_{hair}; \{\vec Q_i\})\,.
\ee
The $n$th term in the sum represents 
the contribution 
from an 
$n$-centered black hole, $\vec Q_i$ denotes 
the charge carried
by the $i$-th center and $\vec Q_{hair}$
denotes the charges carried by the hair 
modes.\footnote{In this section we shall use $\vec Q$ to
denote all the electric and all the magnetic charges, as well
as the angular momentum.}  
In principle, $d_{hair}$ can be calculated by 
explicitly identifying and
quantizing the hair modes.
On the other hand, $d_{hor}(\vec Q_i)$ 
for each center can be computed using
the quantum entropy function formalism described in \S\ref{sqef}.

\subsection{Degeneracy to index} \label{sdeg}

As discussed before, on the microscopic side we usually compute an index. 
On the other hand, $d_{hor}$ computes degeneracy.
More generally, eq.\refb{emacro} gives us a general formula for
computing the degeneracy on the macroscopic side. 
Thus this cannot be directly compared with the $B_6$ index
computed on the microscopic side.

We shall now describe a
strategy for using $d_{hor}$ to compute the index 
on the macroscopic side \cite{0903.1477,1009.3226}. 
We shall illustrate this for the helicity 
trace $B_{n}$ for
four dimensional black holes,
but it can be generalized to five dimensional black
holes as well \cite{0908.3402}. For a black hole that
breaks $2k$ supercharges, we had defined
\be 
\label{bn}
B_{k} = {1\over k!} \, Tr \lbrace(-1)^{2h} (2h)^{k} \rbrace\,,
\ee
where $h$ is the third component of angular momentum in the rest frame.
Since the total contribution to $h$ can be regarded as a sum of the
contributions from the horizon and the hair degrees of freedom,
we can express $B_k$ as
\be  
B_{k;macro} =     
{1\over k!} \, Tr \lbrace (-1)^{2h_{hor} + 2h_{hair}}
(2h_{hor} + 2h_{hair})^k \rbrace \,,
\ee
where $h_{hor}$ and $h_{hair}$ denote the contribution to
$h$ from the horizon and the hair degrees of freedom respectively.

Now, typically all the fermion zero modes   
associated with the broken
supersymmetries are hair degrees of freedom, since 
we can generate these zero mode deformations by 
supersymmetry transformation parameters which go to constant
at infinity and vanish below a certain radius.
Thus the hair modes
contain $2k$ fermion zero modes, and in order that the trace
over these zero modes do not make the whole trace vanish,
we need an insertion  of
$(2h_{hair})^k$ into the trace. 
In other words, if we expand the $(2h_{hor} + 2h_{hair})^k$
factor in a binomial expansion, then only the  
$(2h_{hair})^k$ term will contribute.
This gives
\be 
 B_{k;macro}    
 = {1\over k!} \, Tr \lbrace (-1)^{2h_{hor} + 2h_{hair}}
(2h_{hair})^k  \rbrace =\sum B_{0;hor}\, B_{k;hair} \,.
\ee
This can be expanded in the spirit of \refb{emacro} as 
\be \label{eindex}
B_{k;macro}(\vec Q)=\sum_n\,
\sum_{\{\vec Q_i\}, \vec Q_{hair}\atop \sum_{i=1}^n 
\vec Q_i+ \vec Q_{hair}=\vec Q} 
\, \left\{\prod_{i=1}^n \, B_{0;hor}(\vec Q_i)\right\}  \,
B_{k;hair}(\vec Q_{hair}; \{\vec Q_i\})\,,
\ee
where now the vector $\vec Q$ no longer contains the
angular momentum.     
A further simplification follows from the fact that in four
dimensions, only the $h_{hor}=0$ black holes are supersymmetric.
This is of course known to be true for a classical black hole, but
more generally it follows from the fact that unbroken supersymmetries,
together with the $SL(2,R)$ isometry of the near horizon geometry,
generate the full $SU(1,1|2)$ supergroup which includes $SU(2)$
as a symmetry group. This implies a spherically symmetric horizon,
and hence zero angular momentum since the partition function on
$AdS_2$ computes the entropy in a fixed angular momentum sector 
(microcanonical ensemble).
Thus $B_{0;hor}= Tr_{hor}(1) = d_{hor}\,,$ and we can express
\refb{eindex} as
\be \label{eindex1}
B_{k;macro}(\vec Q)=\sum_n\,
\sum_{\{\vec Q_i\}, \vec Q_{hair}\atop \sum_{i=1}^n 
\vec Q_i+ \vec Q_{hair}=\vec Q} 
\, \left\{\prod_{i=1}^n \, d_{hor}(\vec Q_i)\right\}  \,
B_{k;hair}(\vec Q_{hair}; \{\vec Q_i\})\,.
\ee

Most of our analysis involves 1/4-BPS black holes in $\NN=4$
supersymmetric string theories in $D=4$ which preserves 4 
out of 16 supersymmetries, \i.e., such a black hole 
configuration breaks 12 supersymmetries. Thus the relevant 
helicity trace index is $B_6$. In these
theories, the contribution from
multi-centered black holes is known to be exponentially 
suppressed \cite{0707.1563,0803.3857,0903.2481}.
Furthermore, for single-centered black holes, 
often the only hair modes
are the fermion zero modes.
In this case, $ \vec Q_{hair}=0$. Furthermore, since for each pair
of fermion zero modes $Tr \lbrace (-1)^F (2h)\rbrace =i$, we have 
$ B_{6;hair}=i^6=-1$.
Thus
\be
B_{6;macro} (\vec Q) = - d_{hor}(\vec Q)\,,     
\ee
up to exponentially suppressed contribution from multi-centered
black holes.
This explains how we can compare the helicity trace index
computed in the microscopic theory with $d_{hor}$ computed
in the macroscopic theory.
Note that
since $ d_{hor}(\vec Q)>0$, we get $ B_{6;macro}<0$.    
This agrees with the explicit microscopic results stated 
above \refb{asym} and below  
\refb{eend}.

\begin{table} {\small
\begin{center}\def\st{\vrule height 3ex width 0ex}
\begin{tabular}{|l|l|l|l|l|l|l|l|l|l|l|} \hline 
$(Q^2,P^2){\backslash} Q.P$ &  -2 
& 0 & 1 & 2 & 3 & 4\st\\[1ex] \hline \hline
(2,2) &   -209304 &  {\bf 50064}
&  {\bf 25353} &  648 & 327 & 0 \st\\[1ex] \hline
(2,4) &  -2023536  & {\bf 1127472}
&  {\bf 561576} & {\bf 50064} & 8376 & -648 \st\\[1ex] \hline
(4,4) & -16620544  &  {\bf 32861184} & {\bf 18458000} &  {\bf 3859456}
&  {\bf 561576} & 12800 \st\\[1ex] \hline
(2,6) &  -15493728 &  {\bf 16491600}
&  {\bf 8533821} & {\bf 1127472} & 130329 & -15600 \st\\[1ex] \hline
(4,6) & -53249700 &  {\bf 632078672} & {\bf 392427528}   & 
{\bf 110910300}  &  
{\bf 18458000} 
&  {\bf 1127472} \st\\[1ex] \hline
(6,6) & 2857656828  &  {\bf 16193130552} & {\bf 11232685725}  & 
{\bf 4173501828}
&  {\bf 920577636} & {\bf 110910300}  \st\\[1ex] \hline
 \hline 
\end{tabular}
\caption{Some results for $-B_6$ in heterotic string
theory on $T^6$ for different values of $Q^2$, $P^2$ and
$Q . P$ in a particular chamber of the moduli space. 
The boldfaced entries are for charges
for which only single centered black holes contribute to the index in
the chamber in which $B_6$ is being computed.
} \label{t1}
\end{center} }
\end{table}

The prediction that $B_{6;macro}$ and
hence $B_{6;micro}$ is negative holds even for finite charges
for single centered black holes. Thus if we take the microscopic results for
the index in some specific chamber of the moduli space and then i) either
focus on the charges for which only single centered black holes contribute
to the index in that chamber, or ii) allow the charge to be arbitrary but explicitly
subtract the contribution from the two centered configurations which could
contribute to the index, then the result for $-B_{6;micro}$ 
must be positive in every case.
This has been verified explicitly for all the CHL models for low values of the
charges\cite{1008.4209}. We have shown in table \ref{t1} the result for $-B_6$
for heterotic string theory on $T^6$ for some combinations of the charges.
The boldfaced entries represent charges for which only single centered black
holes contribute to the index, and as we can see, they are all 
positive.\footnote{Using duality invariance of the theory one can argue that
as long in some given chamber $B_6$ is negative for the subset of charge
vectors for which
only single centered black holes contribute to the index, then
this implies negative $B_6$ for all charge vectors as long as we subtract the
contribution of the multi-centered configurations from the total index.}
The complete proof of the positivity of $-B_{6;micro}$ for all charges
is still awaited.

Finally we would like to mention that a similar proof of the equality of
degeneracy and index also exists for five dimensional black 
holes\cite{1009.3226}.

\sectiono{Applications of quantum entropy function} 
\label{sapp}

Eq.\refb{egauss} shows how Wald's formula applied to 1PI
action can be used to calculate 
some of the subleading corrections to
the black hole entropy, and reproduce the results known from
microscopic computation. Since quantum entropy 
function reduces to the exponential of
Wald entropy in the  classical limit, we expect that 
as long as the quantum corrections generate a local
contribution to the 1PI action, Wald's formula 
applied to 1PI action and quantum entropy function will
give the same results.
In this section, we shall describe how quantum entropy function
can be used to compute some other corrections to the entropy
which could not be calculated by direct use of Wald's formula.

\subsection{Computation of twisted index} \label{stw}

Suppose we have a $\ZZZ_N$ symmetry generated by
$g$ that commutes with all the supersymmetries
of an $\NN=4$ supersymmetric string theory.
We can then define a twisted index:
\be B^g_6 = {1\over 6!} Tr \lbrace(-1)^{2h} (2h)^6 g \rbrace \,.
\ee
In \S\ref{sg1} and \S\ref{sg3}, we described 
the results for such indices
in a wide variety of $\NN=4$ supersymmetric string theories.
We shall now describe how to compute them from the
macroscopic side.

We proceed as in \S\ref{sdeg}.
After separating out the contribution from the hair degrees of
freedom, we see that the relevant quantity associated with the
horizon is
\be 
Tr_{hor} \lbrace(-1)^{2h_{hor}} g\rbrace = Tr_{hor}( g)\,,
\ee
since $h_{hor}=0$ for a supersymmetric black hole.
By following the 
logic of $AdS/CFT$ correspondence, we find
that $d_{hor}$ is now given by the finite part of a twisted
partition function
\be 
Z_g = \left\langle \exp[-i  q_k\ointop d\theta \, A^{(k)}_\theta]
\right\rangle_g\,,
\ee
where the subscript $g$ denotes that in carrying out the path
integral, we are instructed to integrate over field
configurations
with a $ g$-twisted boundary condition on the fields under
$\theta\to\theta+ 2\pi$.
Other than this, the asymptotic boundary conditions must be identical to
that of the attractor geometry since the charges have not changed.

From the Euclidean $AdS_2$ metric given in \refb{eclads}, 
we find that the circle at infinity, parametrized by 
$\theta$, is contractible at
the origin $r=1$.
Thus a $g$-twist under $\theta\to\theta + 2\pi$ 
is not admissible.
Hence we conclude that the $AdS_2\times S^2$ 
geometry is not a valid saddle point of the
path integral.
This however is not the end of the story, since according to the
rules of quantum gravity we must sum over all geometries
and field configurations
keeping fixed the asymptotic boundary conditions. Thus we
should investigate
if there are other 
saddle points which could contribute
to the path integral. To find out possible candidates, 
we must keep in mind the following constraints:
\begin{enumerate}
\item It must have the same asymptotic geometry as the $AdS_2\times S^2$
geometry.
\item It must have a $g$-twist under $\theta\to\theta+2\pi$.
\item It must preserve sufficient amount of supersymmetries so that
integration over the fermion zero modes do not make the integral
vanish \cite{0608021,0905.2686}.
\end{enumerate}
There are indeed such saddle points in the path integral, 
constructed as
follows \cite{0911.1563}:   
\begin{enumerate}
\item Take the original near horizon geometry of the black hole.
\item Take a $\ZZZ_{N}$ orbifold of this
background with $\ZZZ_{N}$ generated by the simultaneous action
of 
\begin{enumerate}
\item $ 2\pi/N$ rotation in $AdS_2$ ($\theta\to\theta
+ {2\pi\over N}$),     
\item $2\pi/N$ rotation in $S^2$
($\phi\to\phi
+ {2\pi\over N}$;     
this is needed for preserving SUSY), and 
\item $ g$. 
\end{enumerate}
\end{enumerate}
To see that this has the same asymptotic geometry as the
attractor geometry, we make a rescaling
\be 
\theta\to \theta/N,  \qquad r\to N\, r\,.
\ee
After this rescaling, the metric takes the form:
\be 
ds^2 = v \left( (r^2 - N^{-2}) d\theta^2 + {dr^2 \over r^2 - N^{-2}}
\right) \,,
\ee
with the orbifold action given by: 
\be 
\theta\to\theta+2\pi\,,\qquad \phi\to\phi+2\pi/N \,,\qquad g\,.
\ee
For large $r$, the metric approaches the 
$AdS_2$ metric.\footnote{In contrast, we note 
that for two dimensional flat spacetime, 
orbifolding not only introduces 
a conical singularity but also changes the asymptotic spacetime.}
The $g$ transformation provides us with the correct boundary
condition under $\theta\to\theta+2\pi$.
The shift along the $\phi$-direction can be regarded as a 
Wilson line, and hence is
an allowed fluctuation in $AdS_2$. In other words, by a
coordinate change $\phi \to \phi + \theta/N\,,$ we can remove
the shift in $\phi$, but this will generate a constant
$g_{\theta\phi}$ at the boundary, which describes a normalizable
mode and hence is an allowed fluctuation.

The classical action associated with this orbifold
can be obtained by dividing the action associated with the
parent geometry by $N$. Thus
the classical action associated with this 
saddle point, after
removing the divergent part proportional to the length of the
boundary, is $S_{wald}/N$. 
As a result, the contribution to the finite part of the    
twisted partition function from
this saddle point is
\be 
Z_g^{finite} \sim \exp\left[S_{wald}/N\right]\,.
\ee
This is exactly what we have found in the microscopic analysis
of the twisted index in \S\ref{sg1} and \S\ref{sg3}.

Note that $\exp\left[S_{wald}/N\right]
<<\exp\left[S_{wald}\right]$. Thus the $\ZZZ_{N}$ quantum numbers 
must be delicately distributed among the microstates of the black 
hole so that a charge of order unity, averaged over 
$\exp\left[S_{wald}\right]$ number of states, gives a 
contribution of order $\exp\left[S_{wald}/N\right]$. In other words, 
there is a large cancellation going on among terms of order 
unity to give this result. Nevertheless we see that black 
holes are able to capture
information about this highly sensitive data.

\subsection{Logarithmic corrections to the black hole entropy}
\label{slog}

As already discussed before,
the effect of integrating out the massive mode
contribution to $Z_{AdS_2}$
can be regarded as a modification of the effective Lagrangian
density, and can be accommodated using Wald's formula.
However, for calculating the one loop contribution due to the
massless modes, we need to compute directly the determinant
of the kinetic operator in the $AdS_2\times S^2$ 
background.

\begin{table} {\small
\begin{center}\def\st{\vrule height 3ex width 0ex}
\begin{tabular}{|l|l|l|l|l|l|l|l|l|l|l|} \hline
The theory  & scaling of charges & logarithmic contribution & microscopic
\st\\[1ex] \hline \hline
$\NN=4$ supersymmetric CHL 
&  ${  Q_i}\sim \Lambda$,
\quad A $\sim\Lambda^2$  &   0 & $\surd$ \st\\[0 ex]
models in $D=4$ and type II on  &&& \st\\[0 ex]
$K3\times T^2$ with $  n_v$ matter multiplet &&& 
\st\\[1ex] \hline
Type II on $T^6$&   ${  Q_i}\sim \Lambda$, ${  A}\sim \Lambda^2$
& $-{  8\, \ln}\, \Lambda$ & $\surd$
\st\\[1ex] 
\hline
BMPV in type IIB on $  T^5/\ZZZ_N$ &  ${  Q_1, Q_5, n}\sim \Lambda$, & 
 ${  -{1\over 4} (n_V-3) \, \ln} \, \Lambda$ & $\surd$
\st\\[0 ex] 
or $  K3\times S^1/\ZZZ_N$ with $  n_V$ vectors & 
${  J}\sim \Lambda^{3/2}$, \quad A $\sim\Lambda^{3/2}$
& & \st\\[0ex]
preserving 16 or 32 supercharges &&& 
\st\\[1ex] \hline
BMPV in type IIB on $T^5/\ZZZ_N$ &  ${  Q_1, Q_5, n}\sim \Lambda$, & 
 ${  -{1\over 4} (n_V+3) \, \ln} \, \Lambda$ & $\surd$
\st\\[0 ex] 
or $  K3\times S^1/\ZZZ_N$ with $  n_V$ vectors & ${  J=0}$, 
\quad A $\sim\Lambda^{3/2}$
& & \st\\[0ex]
preserving 16 or 32 supercharges &&& 
\st\\[1ex] \hline
$\NN=2$ supersymmetric theories
 &   ${ 
Q_i}\sim\Lambda$, \quad A $\sim\Lambda^2$
&  ${  {1\over 6} (23 + n_H - n_V)\, \ln}\,  \Lambda$
& ?\st\\[0 ex] in $D=4$ with $  n_V$ vector and &&&
\st\\[0 ex]   $  n_H$  hyper multiplets & & &
\st\\[1ex] \hline
$\NN=6$ supersymmetric theory
&  ${  Q_i}\sim \Lambda$, \quad A $\sim\Lambda^2$
 & ${  -4\, \ln} \, \Lambda$ & ?  \st\\[0ex]
 in $D=4$ &&& \st\\[1ex] \hline
$\NN=5$ supersymmetric theory &  ${  Q_i}\sim \Lambda$, \quad A $\sim\Lambda^2$
 & ${  -2\, \ln} \, \Lambda$ & ? \st\\[0ex]
 in $D=4$ &&&
\st\\[1ex] \hline
$\NN=3$ supersymmetric theory in &  ${  Q_i}\sim \Lambda$, 
\quad A $\sim\Lambda^2$ & 
 ${  2\, \ln} \, \Lambda$ & ? \st\\[0 ex]
$D=4$ with $  n_v$ matter multiplets &&& \st\\[1ex] \hline\hline
\end{tabular}
\end{center}
\caption{A table showing the macroscopic predictions for the logarithmic
corrections to extremal black hole entropy in a wide class of string theories and
the status of their comparison with the microscopic results. In the last column
a $\surd$ indicates that the microscopic results are available and agree with the
macroscopic prediction while a ? indicates that the microscopic results are
not yet available. The first column describes the theory and the black hole
under consideration, the second column describes the scaling of the various
charges under which the logarithmic correction is computed and also how
the area $A$ of the event horizon scales under these scalings of the charges.
The third column describes the macroscopic results for the 
logarithmic correction to the entropy under 
this scaling. Unless labelled otherwise, $Q_i$ in the second column stands
for all the electric and magnetic charges of the black hole. For BMPV black
holes $Q_1$, $Q_5$, $n$ and $J$ stand respectively for the D1-brane charge,
D5-brane charge, Kaluza-Klein momentum
and the angular momentum.} \label{t2} }
\end{table}

Let us consider an example where we have a massless scalar field 
with the standard kinetic term in the near horizon
$AdS_2\times S^2$ background for a spherically
symmetric extremal black hole in $D=4$.
All the eigenvalues and eigenfunctions of $\square$ on
$AdS_2\times S^2$ can be found explicitly, which can then be used to
compute $\det\,\, \square$, 
and
hence the one loop contribution to $ Z_{AdS_2}$. The result for the
contribution to $ \ln d_{hor}$ from this massless scalar is of the 
form\footnote{A different approach
to computing logarithmic
corrections to extremal black hole entropy can be found
in \cite{9709064}.} \cite{1005.3044}:
\be
-{1\over 180} \, \ln A \,.
\ee
For black holes in supergravity/superstring theory, the kinetic
operator for fluctuations around the near horizon geometry 
mixes scalars, vectors and tensors.
Thus one needs to diagonalize the kinetic operator, find the
determinant, and then compute its contribution to
$Z_{AdS_2}$ and hence $d_{hor}$.
This has been achieved for BPS black holes in $\NN=8,6,5,4,3,2$
supersymmetric theories in four 
dimensions\cite{1005.3044,1106.0080,1108.3842,1109.0444}
and for BMPV black holes
in five dimensions\cite{1109.3706} 
and in whichever case the microscopic results
are available, {\it e.g.} for $\NN=4$ and 8\cite{0908.0039} 
supersymmetric theories in
four dimensions and BMPV black holes in five dimensional
theories with 16 or
32 supersymmetries, the 
macroscopic results are in perfect agreement with the microscopic
results. 
The situation has been summarized in table \ref{t2}.

\subsection{Other applications} \label{sother}

Quantum entropy function has also been used to explain several
other features of the microscopic formula. For example, we see from
the microscopic formula
\refb{epropr} that for charge vectors $(Q,P)$ 
with $r(Q,P)>1$, there are additional
contributions to the $B_6$ index whose leading term takes the
form $\exp\left(\pi\sqrt{Q^2P^2-(Q\cdot P)^2}/s\right)$, where $s$ is
a factor of $r$. It turns out that precisely for $r(Q,P)>1$, the functional
integral for $Z_{AdS_2}$ receives extra contribution from
saddle points obtained by taking a freely acting
$\ZZZ_s$ quotient -- for $s|r$ --  of the original
near horizon geometry. The leading semi-classical contribution from
such a saddle point is given  by $\exp(S_{wald}/s)=
 \exp\left(\pi\sqrt{Q^2P^2-(Q\cdot P)^2}/s\right)$, precisely in
 agreement with the microscopic results\cite{0903.1477,0908.0039}. 
 
For $r=1$, the result for $B_6$ for large charges takes the form
of a sum of the contributions from different poles. The leading
asymptotic expansion comes from a specific pole and is given by
\refb{pffourier2}. It turns out that the contributions from the other poles
have the leading term of the form $\exp\left(\pi\sqrt{Q^2P^2-(Q\cdot P)^2}/
N\right)$, for $N\in \ZZZ$, $N>1$. 
On the other hand,    
$Z_{AdS_2}$ receives contribution from, besides the original near
horizon geometry, its $\ZZZ_N$ orbifolds which do not change the
boundary condition at infinity. The leading semiclassical
contribution from these saddle points is given by 
$\exp\left(\pi\sqrt{Q^2P^2-(Q\cdot P)^2}/
N\right)$, precisely in correspondence with the leading contribution from the
subleading poles in the microscopic 
formula\cite{0810.3472,0904.4253}. 

Eventually we hope to reproduce the complete asymptotic
expansion of the microscopic result for $\ln|B_6|$ (or $\ln |B_{14}|$ 
for type II on $T^6$)
from the 
string theory path integral over $ AdS_2$.
One possible tool one could use for this
is the localization of the path integral to a finite
dimensional subspace using 
supersymmetry.
This has been pursued to some extent in \cite{0905.2686}
and has been further developed in \cite{1012.0265,1111.1161}.
In particular \cite{1012.0265,1111.1161} managed to localize the
path integral over vector multiplet moduli fields by expressing the
full supergravity path integral as an integral over various
supersmultiplets of an $\NN=2$ supersymmetric theory. 
However
the path integral over the hyper, gravity and gravitino (for $\NN>2$
supersymmetric theories) multiplets still remains to be understood.
Despite this the analysis of \cite{1012.0265,1111.1161} already
gives some encouraging results. 
In particular assuming that the integration over the other fields do not
contribute to the final result, \cite{1111.1161} was able to reproduce the
asymptotic expansion of the result \refb{eb14} for black holes in type IIB
string theory on $T^6$.

\sectiono{Discussion}

All these results provide us with the 
`experimental verification' of the theory of extremal black
holes, based on Wald's formula and
$AdS_2/CFT_1$ correspondence.
The results described here show that
quantum gravity in the near horizon geometry 
contains detailed information about not only
the total number of microstates but also finer details ({\it e.g.}
the $\ZZZ_N$ quantum numbers carried by the microstates).
Thus, at least for extremal black holes, there seems to be an exact
duality between
\be
Gravity \,\,\,description\,\,\, \Leftrightarrow\,\,\, 
Microscopic\,\,\, description\,.
\ee
The gravity description contains as much information
as the microscopic description,
but in a quite different way.

It is clear from our discussions that whereas the 
$\alpha'$-corrections are well-understood through 
Wald's formalism, we need to understand the $g_s$ 
corrections better. The quantum entropy function 
formalism provides us with a tool for investigations in that 
direction but this requires carrying out the functional integral over
the string fields in the near horizon geometry of the black hole. 
In this process of evaluating the path integral over the near horizon
geometry, we hope to learn not only about black holes
but also about string theory, {\it e.g.} the rules for carrying out
path integral over string fields.

Another useful direction of study is the generalization of these
results to $\NN=2$ supersymmetric string theories.
Some attempts at generalizing the microscopic results 
of \S\ref{stwo} in
special $\NN=2$ supersymmetric string theories can be
found in \cite{0711.1971,0810.1233,0905.4115}, while a general
asymptotic formula can be found in \cite{0702146}.

{\bf Acknowledgement:}  
We would like to thank Nabamita Banerjee, Shamik Banerjee, Atish
Dabholkar, Justin David,
Suvankar Dutta, Dileep Jatkar, Joao Gomes, Rajesh Gopakumar,
Rajesh Gupta and
Sameer Murthy for collaboration and/or 
useful discussions. A.S. would like to thank the Perimeter institute, Canada and the Simons Center at Stony Brook
for hospitality 
during the preparation of this manuscript.
The work of I.M. was supported in part by DAE
project 11-R\&D-HRI-5.02-0304 and the grant from the 
Chaires Internationales de Recherche Blaise Pascal of A.S.
The work of A.S. 
was supported in part by the J. C. Bose fellowship of the Department of
Science and Technology, India, the DAE
project 11-R\&D-HRI-5.02-0304, and by
the Chaires Internationales de Recherche Blaise Pascal, France.


\small
\baselineskip 12pt


\begin{thebibliography}{99}

\bibitem{r1}
  S.~W.~Hawking,
  Phys.\ Rev.\ Lett.\  {\bf 26}, 1344 (1971).

\bibitem{r2}
  J.~D.~Bekenstein,
  Phys.\ Rev.\  D {\bf 7}, 2333 (1973).

\bibitem{r3}
  J.~D.~Bekenstein,
  Phys.\ Rev.\  D {\bf 9}, 3292 (1974).

\bibitem{r4}
  S.~W.~Hawking,
  Nature {\bf 248}, 30 (1974).


\bibitem{9504147}
  A.~Sen,
  Mod.\ Phys.\ Lett.\  A {\bf 10}, 2081 (1995)
  [arXiv:hep-th/9504147].

\bibitem{9601029}
  A.~Strominger and C.~Vafa,
  Phys.\ Lett.\  B {\bf 379}, 99 (1996)
  [arXiv:hep-th/9601029].

\bibitem{0810.4525}
  S.~D.~Mathur,
  arXiv:0810.4525 [hep-th], and references therein.

\bibitem{9607026}
R.~Dijkgraaf, E.~P.~Verlinde and H.~L.~Verlinde,
Nucl.\ Phys.\ B {\bf 484}, 543 (1997)
[arXiv:hep-th/9607026].

\bibitem{0412287}
  G.~Lopes Cardoso, B.~de Wit, J.~Kappeli and T.~Mohaupt,
  JHEP {\bf 0412}, 075 (2004)
  [arXiv:hep-th/0412287].

\bibitem{0505094}
D.~Shih, A.~Strominger and X.~Yin,
JHEP {\bf 0610}, 087 (2006)
  [arXiv:hep-th/0505094].

\bibitem{0506151}
  D.~Shih, A.~Strominger and X.~Yin,
  JHEP {\bf 0606}, 037 (2006)
  [arXiv:hep-th/0506151].

\bibitem{0506249}
D.~Gaiotto,
arXiv:hep-th/0506249.

\bibitem{0508174}
  D.~Shih and X.~Yin,
  JHEP {\bf 0604}, 034 (2006)
  [arXiv:hep-th/0508174].

\bibitem{0510147}
  D.~P.~Jatkar and A.~Sen,
  JHEP {\bf 0604}, 018 (2006)
  [arXiv:hep-th/0510147].

\bibitem{0602254}
  J.~R.~David, D.~P.~Jatkar and A.~Sen,
  JHEP {\bf 0606}, 064 (2006)
  [arXiv:hep-th/0602254].


\bibitem{0603066}
  A.~Dabholkar and S.~Nampuri,
  JHEP {\bf 0711}, 077 (2007)
  [arXiv:hep-th/0603066].

\bibitem{0605210}
  J.~R.~David and A.~Sen,
  JHEP {\bf 0611}, 072 (2006)
  [arXiv:hep-th/0605210].

\bibitem{0607155}
  J.~R.~David, D.~P.~Jatkar and A.~Sen,
  JHEP {\bf 0611}, 073 (2006)
  [arXiv:hep-th/0607155].

\bibitem{0609109}
  J.~R.~David, D.~P.~Jatkar and A.~Sen,
  JHEP {\bf 0701}, 016 (2007)
  [arXiv:hep-th/0609109].

\bibitem{0612011}
  A.~Dabholkar and D.~Gaiotto,
  JHEP {\bf 0712}, 087 (2007)
  [arXiv:hep-th/0612011]

\bibitem{0702141}
  A.~Sen,
  JHEP {\bf 0705}, 039 (2007)
  [arXiv:hep-th/0702141].

\bibitem{0702150}
  A.~Dabholkar, D.~Gaiotto and S.~Nampuri,
  JHEP {\bf 0801}, 023 (2008)
  [arXiv:hep-th/0702150].

\bibitem{0705.1433}
  N.~Banerjee, D.~P.~Jatkar and A.~Sen,
  JHEP {\bf 0707}, 024 (2007)
  [arXiv:0705.1433 [hep-th]].

\bibitem{0705.3874}
  A.~Sen,
  JHEP {\bf 0709}, 045 (2007)
  [arXiv:0705.3874 [hep-th]].

\bibitem{0706.2363}
  M.~C.~N.~Cheng and E.~Verlinde,
  JHEP {\bf 0709}, 070 (2007)
  [arXiv:0706.2363 [hep-th]].

\bibitem{0707.1563}
  A.~Sen,
  JHEP {\bf 0710}, 059 (2007)
  [arXiv:0707.1563 [hep-th]].

\bibitem{0707.3035}
  A.~Mukherjee, S.~Mukhi and R.~Nigam,
  JHEP {\bf 0710}, 037 (2007)
  [arXiv:0707.3035 [hep-th]].

\bibitem{0708.1270}
  A.~Sen,
Gen.\ Rel.\ Grav.\  {\bf 40}, 2249 (2008)
  [arXiv:0708.1270 [hep-th]].

\bibitem{0708.3715}
  A.~Sen,
  JHEP {\bf 0712}, 019 (2007)
  [arXiv:0708.3715 [hep-th]].

\bibitem{0710.4533}
  A.~Mukherjee, S.~Mukhi and R.~Nigam,
  Mod.\ Phys.\ Lett.\  A {\bf 24}, 1507 (2009)
  [arXiv:0710.4533 [hep-th]].

\bibitem{0712.0043}
  S.~Banerjee and A.~Sen,
  JHEP {\bf 0803}, 022 (2008)
  [arXiv:0712.0043 [hep-th]].

\bibitem{0801.0149}
  S.~Banerjee and A.~Sen,
  JHEP {\bf 0804}, 012 (2008)
  [arXiv:0801.0149 [hep-th]].


\bibitem{0802.0544}
  S.~Banerjee, A.~Sen and Y.~K.~Srivastava,
  JHEP {\bf 0805}, 101 (2008)
  [arXiv:0802.0544 [hep-th]].

\bibitem{0802.0761}
  A.~Dabholkar, K.~Narayan and S.~Nampuri,
  JHEP {\bf 0803}, 026 (2008)
  [arXiv:0802.0761 [hep-th]].

\bibitem{0802.1556}
  S.~Banerjee, A.~Sen and Y.~K.~Srivastava,
  arXiv:0802.1556 [hep-th].

\bibitem{0803.1014}
  A.~Sen,
  JHEP {\bf 0807}, 118 (2008)
  [arXiv:0803.1014 [hep-th]].

\bibitem{0803.2692}
  A.~Dabholkar, J.~Gomes and S.~Murthy,
  JHEP {\bf 0805}, 098 (2008)
  [arXiv:0802.1556 [hep-th]].

\bibitem{0803.3857}
  A.~Sen,
  JHEP {\bf 0807}, 078 (2008)
  [arXiv:0803.3857 [hep-th]].

\bibitem{0806.2337}
  M.~C.~N.~Cheng and E.~P.~Verlinde,
  SIGMA {\bf 4}, 068 (2008)
  [arXiv:0806.2337 [hep-th]].

\bibitem{0807.0237}
  A.~Castro and S.~Murthy,
  JHEP {\bf 0906}, 024 (2009)
  [arXiv:0807.0237 [hep-th]].

\bibitem{0807.1314}
  N.~Banerjee,
  Phys.\ Rev.\  D {\bf 79}, 081501 (2009)
  [arXiv:0807.1314 [hep-th]].

\bibitem{0807.4451}
  S.~Govindarajan and K.~Gopala Krishna,
  JHEP {\bf 0904}, 032 (2009)
  [arXiv:0807.4451 [hep-th]].

\bibitem{0808.1746}
  S.~Banerjee, A.~Sen and Y.~K.~Srivastava,
  JHEP {\bf 0903}, 151 (2009)
  [arXiv:0808.1746 [hep-th]].

\bibitem{0809.1157}
  S.~Mukhi and R.~Nigam,
  JHEP {\bf 0812}, 056 (2008)
  [arXiv:0809.1157 [hep-th]].

\bibitem{0809.4258}
  M.~C.~N.~Cheng and A.~Dabholkar,
  Commun.\ Num.\ Theor.\ Phys.\  {\bf 3}, 59 (2009)
  [arXiv:0809.4258 [hep-th]].

\bibitem{0810.3472}
  N.~Banerjee, D.~P.~Jatkar and A.~Sen,
  JHEP {\bf 0905}, 121 (2009)
  [arXiv:0810.3472 [hep-th]].

\bibitem{0901.1758}
  M.~C.~N.~Cheng and L.~Hollands,
  JHEP {\bf 0904}, 067 (2009)
  [arXiv:0901.1758 [hep-th]].

\bibitem{0903.2481}
  A.~Dabholkar, M.~Guica, S.~Murthy and S.~Nampuri,
  JHEP {\bf 1006}, 007 (2010)
  [arXiv:0903.2481 [hep-th]].

\bibitem{0907.1410}
  S.~Govindarajan and K.~Gopala Krishna,
  JHEP {\bf 1005}, 014 (2010)
  [arXiv:0907.1410 [hep-th]].

\bibitem{0911.0586}
  A.~Dabholkar and J.~Gomes,
  JHEP {\bf 1003}, 128 (2010)
  [arXiv:0911.0586 [hep-th]].

\bibitem{0911.1563}
  A.~Sen,
  JHEP {\bf 1005}, 028 (2010)
  [arXiv:0911.1563 [hep-th]].

\bibitem{1002.3857}
  A.~Sen,
  arXiv:1002.3857 [hep-th].

\bibitem{1006.3472}
  S.~Govindarajan,
  arXiv:1006.3472 [hep-th].

\bibitem{9611205}
  C.~Bachas and E.~Kiritsis,
  Nucl.\ Phys.\ Proc.\ Suppl.\  {\bf 55B}, 194 (1997)
  [arXiv:hep-th/9611205].

\bibitem{9708062}
  A.~Gregori, E.~Kiritsis, C.~Kounnas, N.~A.~Obers, P.~M.~Petropoulos and B.~Pioline,
  Nucl.\ Phys.\  B {\bf 510}, 423 (1998)
  [arXiv:hep-th/9708062].

\bibitem{0010222}
F.~Denef,
arXiv:hep-th/0010222.

\bibitem{0101135}
  F.~Denef, B.~R.~Greene and M.~Raugas,
  JHEP {\bf 0105}, 012 (2001)
  [arXiv:hep-th/0101135].

\bibitem{0206072}
  F.~Denef,
  JHEP {\bf 0210}, 023 (2002)
  [arXiv:hep-th/0206072].

\bibitem{0304094}
  B.~Bates and F.~Denef,
  arXiv:hep-th/0304094.

\bibitem{0702146}
  F.~Denef and G.~W.~Moore,
  arXiv:hep-th/0702146.

  \bibitem{9602065}
  J.~C.~Breckenridge, R.~C.~Myers, A.~W.~Peet and C.~Vafa,
  Phys.\ Lett.\ B {\bf 391}, 93 (1997)
  [arXiv:hep-th/9602065].

\bibitem{0503217}
  D.~Gaiotto, A.~Strominger and X.~Yin,
  JHEP {\bf 0602}, 024 (2006)
  [arXiv:hep-th/0503217].


\bibitem{9603078}
  J.~C.~Breckenridge, D.~A.~Lowe, R.~C.~Myers, A.~W.~Peet, A.~Strominger 
and C.~Vafa,
  Phys.\ Lett.\  B {\bf 381}, 423 (1996)
  [arXiv:hep-th/9603078].


\bibitem{brill}
D.~Brill, Phys. Rev. B133 (1964) 845.

  \bibitem{pope}
 C.~N.~Pope,
  Nucl.\ Phys.\ B {\bf 141}, 432 (1978).

\bibitem{9608096}
  R.~Dijkgraaf, G.~W.~Moore, E.~P.~Verlinde and H.~L.~Verlinde,
  Commun.\ Math.\ Phys.\  {\bf 185}, 197 (1997)
  [arXiv:hep-th/9608096].


\bibitem{igusa1}
J. Igusa,  
Amer. J. Math. 84 (1962) 175200.

\bibitem{igusa2}
J. Igusa,  
Amer. J. Math. 86 (1962) 392412.

\bibitem{0601108}
  G.~Lopes Cardoso, B.~de Wit, J.~Kappeli and T.~Mohaupt,
  JHEP {\bf 0603}, 074 (2006)
  [arXiv:hep-th/0601108].


 \bibitem{9507090}
  M.~Cvetic and D.~Youm,
  Phys.\ Rev.\ D {\bf 53}, 584 (1996)
  [arXiv:hep-th/9507090].
 
 \bibitem{9508094}
  M.~J.~Duff, J.~T.~Liu and J.~Rahmfeld,
  Nucl.\ Phys.\ B {\bf 459}, 125 (1996)
  [arXiv:hep-th/9508094].

\bibitem{9512031}  
M.~Cvetic and A.~A.~Tseytlin,
Phys.\ Rev.\\uffff D {\bf 53}, 5619 (1996)
[Erratum-ibid.\\uffff D {\bf 55}, 3907 (1997)]
[arXiv:hep-th/9512031].

\bibitem{9508144}
  S.~Chaudhuri and D.~A.~Lowe,
  Nucl.\ Phys.\  B {\bf 459}, 113 (1996)
  [arXiv:hep-th/9508144].

\bibitem{9508154}
  P.~S.~Aspinwall,
  Nucl.\ Phys.\ Proc.\ Suppl.\  {\bf 46}, 30 (1996)
  [arXiv:hep-th/9508154].

\bibitem{borcherds}
R.~Borcherds,  
Invent. Math. {\bf 120} (1995) 161.

\bibitem{9504006}
V.~A.~Gritsenko and V.~V.~Nikulin, 
Amer.\ J.\ Math.\ {\bf 119}, 181 (1997)
[arXiv:alg-geom/9504006].

\bibitem{ibu1}
T. Ibukiyama,  
Int. J. Math 2(1) (1991) 1735. 

\bibitem{ibu2}
S. Hayashida and T. Ibukiyama,  
Journal of Kyoto Univ 45 (2005). 

\bibitem{ibu3}
H. Aoki and T. Ibukiyama, 
International Journal of Mathematics 16 (2005) 
249279. 

\bibitem{eichler}
M. Eichler and D. Zagier,  The theory of jacobi forms, Birkhauser (1985). 

\bibitem{skor}
N.~-P.~Skoruppa, Mathematics of Computation, {\bf 58},
381 (1992).

\bibitem{rama}
M.~Manickam, B.~Ramakrishnan and T.~C.~Vasudevan,
Manuscripta Math.,  {\bf 81}, 161 (1993).

\bibitem{9505054}
  S.~Chaudhuri, G.~Hockney and J.~D.~Lykken,
  Phys.\ Rev.\ Lett.\  {\bf 75}, 2264 (1995)
  [arXiv:hep-th/9505054].

\bibitem{9506048}
  S.~Chaudhuri and J.~Polchinski,
  Phys.\ Rev.\  D {\bf 52}, 7168 (1995)
  [arXiv:hep-th/9506048].

\bibitem{0804.0651}
  A.~Sen,
  JHEP {\bf 0808}, 037 (2008)
  [arXiv:0804.0651 [hep-th]].

\bibitem{0908.0039}
  A.~Sen,
  JHEP {\bf 1002}, 090 (2010)
  [arXiv:0908.0039 [hep-th]].

\bibitem{1109.3706}
  A.~Sen,
  ``Logarithmic Corrections to Rotating Extremal 
  Black Hole Entropy in Four and Five Dimensions,''
  [arXiv:1109.3706 [hep-th]].

\bibitem{1009.3226}
A.~Dabholkar, J.~Gomes, S.~Murthy and A.~Sen,
  JHEP {\bf 1104}, 034 (2011)
  [arXiv:1009.3226 [hep-th]].



\bibitem{9707015}
  O.~B.~Zaslavsky,
  Phys.\ Rev.\  D {\bf 56}, 2188 (1997)
  [Erratum-ibid.\  D {\bf 59}, 069901 (1999)]
  [arXiv:gr-qc/9707015].

\bibitem{9709064}
  R.~B.~Mann and S.~N.~Solodukhin,
  Nucl.\ Phys.\  B {\bf 523}, 293 (1998)
  [arXiv:hep-th/9709064].

\bibitem{9812073}
  J.~M.~Maldacena, J.~Michelson and A.~Strominger,
  JHEP {\bf 9902}, 011 (1999)
  [arXiv:hep-th/9812073].

\bibitem{0705.4214}
  H.~K.~Kunduri, J.~Lucietti and H.~S.~Reall,
  Class.\ Quant.\ Grav.\  {\bf 24}, 4169 (2007)
  [arXiv:0705.4214 [hep-th]].

\bibitem{0803.2998}
  P.~Figueras, H.~K.~Kunduri, J.~Lucietti and M.~Rangamani,
  Phys.\ Rev.\  D {\bf 78}, 044042 (2008)
  [arXiv:0803.2998 [hep-th]].

\bibitem{9307038}
  R.~M.~Wald,
  Phys.\ Rev.\ D {\bf 48}, 3427 (1993)
  [arXiv:gr-qc/9307038].

\bibitem{9312023}
  T.~Jacobson, G.~Kang and R.~C.~Myers,
  Phys.\ Rev.\ D {\bf 49}, 6587 (1994)
  [arXiv:gr-qc/9312023].

\bibitem{9403028}
  V.~Iyer and R.~M.~Wald,
  Phys.\ Rev.\ D {\bf 50}, 846 (1994)
  [arXiv:gr-qc/9403028].

\bibitem{9502009}
  T.~Jacobson, G.~Kang and R.~C.~Myers,
  arXiv:gr-qc/9502009.

\bibitem{0506177}
  A.~Sen,
  JHEP {\bf 0509}, 038 (2005)
  [arXiv:hep-th/0506177].

\bibitem{0508042}
  A.~Sen,
  JHEP {\bf 0603}, 008 (2006)
  [arXiv:hep-th/0508042].



\bibitem{0606244}
  D.~Astefanesei, K.~Goldstein, R.~P.~Jena, A.~Sen and S.~P.~Trivedi,
  JHEP {\bf 0610}, 058 (2006)
  [arXiv:hep-th/0606244].

\bibitem{9508072}
  S.~Ferrara, R.~Kallosh and A.~Strominger,
  Phys.\ Rev.\ D {\bf 52}, 5412 (1995)
  [arXiv:hep-th/9508072].

\bibitem{9602111}
  A.~Strominger,
  Phys.\ Lett.\ B {\bf 383}, 39 (1996)
  [arXiv:hep-th/9602111].

\bibitem{9602136}
  S.~Ferrara and R.~Kallosh,
  Phys.\ Rev.\ D {\bf 54}, 1514 (1996)
  [arXiv:hep-th/9602136].

\bibitem{0507096}
  K.~Goldstein, N.~Iizuka, R.~P.~Jena and S.~P.~Trivedi,
  Phys.\ Rev.\ D {\bf 72}, 124021 (2005)
  [arXiv:hep-th/0507096].

\bibitem{9711053}
J.~M.~Maldacena, A.~Strominger and E.~Witten,
JHEP {\bf 9712}, 002 (1997)
[arXiv:hep-th/9711053].

\bibitem{9801081} 
K.~Behrndt, G.~Lopes Cardoso, B.~de Wit, 
D.~Lust, T.~Mohaupt and W.~A.~Sabra,
Phys.\ Lett.\ B {\bf 429}, 289 (1998) [arXiv:hep-th/9801081].

\bibitem{9812082}
G.~Lopes Cardoso, B.~de Wit and T.~Mohaupt,
Phys.\ Lett.\ B {\bf 451}, 309 (1999)
[arXiv:hep-th/9812082].

\bibitem{9904005}
G.~Lopes Cardoso, B.~de Wit and T.~Mohaupt,
Fortsch.\ Phys.\  {\bf 48}, 49 (2000)
[arXiv:hep-th/9904005].

\bibitem{9906094}
  G.~Lopes Cardoso, B.~de Wit and T.~Mohaupt,
  Nucl.\ Phys.\  B {\bf 567}, 87 (2000)
  [arXiv:hep-th/9906094].

\bibitem{9910179}
G.~Lopes Cardoso, B.~de Wit and T.~Mohaupt,
Class.\ Quant.\ Grav.\  {\bf 17}, 1007 (2000)
[arXiv:hep-th/9910179].

\bibitem{0007195}
  T.~Mohaupt,
  Fortsch.\ Phys.\  {\bf 49}, 3 (2001)
  [arXiv:hep-th/0007195].

\bibitem{0009234}
G.~Lopes Cardoso, B.~de Wit, J.~Kappeli and T.~Mohaupt,
interactions,''
JHEP {\bf 0012}, 019 (2000)
[arXiv:hep-th/0009234].

\bibitem{0012232}
G.~L.~Cardoso, B.~de Wit, J.~Kappeli and T.~Mohaupt,
with
R**2-interactions,''
Fortsch.\ Phys.\  {\bf 49}, 557 (2001)
[arXiv:hep-th/0012232].

\bibitem{0405146}
  H.~Ooguri, A.~Strominger and C.~Vafa,
  Phys.\ Rev.\  D {\bf 70}, 106007 (2004)
  [arXiv:hep-th/0405146].
  
\bibitem{9610237}
  J.~A.~Harvey and G.~W.~Moore,
  Phys.\ Rev.\  D {\bf 57}, 2323 (1998)
  [arXiv:hep-th/9610237].

\bibitem{0603149}
  B.~Sahoo and A.~Sen,
  JHEP {\bf 0609}, 029 (2006)
  [arXiv:hep-th/0603149].

\bibitem{0612225}
  G.~L.~Cardoso, B.~de Wit and S.~Mahapatra,
  JHEP {\bf 0703}, 085 (2007)
  [arXiv:hep-th/0612225].

\bibitem{0808.2627}
  G.~L.~Cardoso, B.~de Wit and S.~Mahapatra,
  JHEP {\bf 0902}, 006 (2009)
  [arXiv:0808.2627 [hep-th]].

\bibitem{0809.3304}
  A.~Sen,
  Int.\ J.\ Mod.\ Phys.\  A {\bf 24}, 4225 (2009)
  [arXiv:0809.3304 [hep-th]].

\bibitem{0903.1477}
  A.~Sen,
  JHEP {\bf 0908}, 068 (2009)
  [arXiv:0903.1477 [hep-th]].

\bibitem{0106112}
  J.~M.~Maldacena,
  JHEP {\bf 0304}, 021 (2003)
  [arXiv:hep-th/0106112].

\bibitem{0710.2956}
  T.~Azeyanagi, T.~Nishioka and T.~Takayanagi,
  Phys.\ Rev.\  D {\bf 77}, 064005 (2008)
  [arXiv:0710.2956 [hep-th]].

\bibitem{0901.0359}
  N.~Banerjee, I.~Mandal and A.~Sen,
  JHEP {\bf 0907}, 091 (2009)
  [arXiv:0901.0359 [hep-th]].

\bibitem{0907.0593}
  D.~P.~Jatkar, A.~Sen and Y.~K.~Srivastava,
  JHEP {\bf 1002}, 038 (2010)
  [arXiv:0907.0593 [hep-th]].

\bibitem{0908.3402}
  A.~Sen,
  JHEP {\bf 1005}, 097 (2010)
  [arXiv:0908.3402 [hep-th]].

\bibitem{1008.4209}
  A.~Sen,
  [arXiv:1008.4209 [hep-th]].


\bibitem{0608021}
  C.~Beasley, D.~Gaiotto, M.~Guica, L.~Huang, A.~Strominger and X.~Yin,
  arXiv:hep-th/0608021.

\bibitem{0905.2686}
  N.~Banerjee, S.~Banerjee, R.~K.~Gupta, I.~Mandal and A.~Sen,
  JHEP {\bf 1002}, 091 (2010)
  [arXiv:0905.2686 [hep-th]].

\bibitem{1005.3044}
  S.~Banerjee, R.~K.~Gupta and A.~Sen,
  arXiv:1005.3044 [hep-th].

\bibitem{1106.0080}
  S.~Banerjee, R.~K.~Gupta, I.~Mandal, A.~Sen,
  ``Logarithmic Corrections to N=4 and N=8 Black Hole Entropy: A One Loop Test of Quantum Gravity,''
  [arXiv:1106.0080 [hep-th]].
  
\bibitem{1108.3842}
  A.~Sen,
  ``Logarithmic Corrections to N=2 Black Hole Entropy: 
  An Infrared Window into the Microstates,''
  [arXiv:1108.3842 [hep-th]].

\bibitem{1109.0444}
S.~Ferrara and A.~Marrani,  
``Generalized Mirror Symmetry and Quantum Black Hole Entropy'',
[arXiv:1109.0444 [hep-th]].

\bibitem{0904.4253}
  S.~Murthy and B.~Pioline,
  JHEP {\bf 0909}, 022 (2009)
  [arXiv:0904.4253 [hep-th]].

\bibitem{1012.0265}
  A.~Dabholkar, J.~Gomes and S.~Murthy,
  JHEP {\bf 1106} (2011) 019
  [arXiv:1012.0265 [hep-th]].

\bibitem{1111.1161}
  A.~Dabholkar, J.~Gomes and S.~Murthy,
  arXiv:1111.1161 [hep-th].

\bibitem{0711.1971}
  J.~R.~David,
  JHEP {\bf 0802}, 025 (2008)
  [arXiv:0711.1971 [hep-th]].

\bibitem{0810.1233}
  G.~L.~Cardoso, J.~R.~David, B.~de Wit and S.~Mahapatra,
  JHEP {\bf 0812}, 086 (2008)
  [arXiv:0810.1233 [hep-th]].

\bibitem{0905.4115}
  J.~R.~David,
  JHEP {\bf 0908}, 054 (2009)
  [arXiv:0905.4115 [hep-th]].

\end{thebibliography}
\end{document}